\documentclass[aps,pre,reprint
]{revtex4-2}
\usepackage{eurosym}
\usepackage{graphicx}
\usepackage{amsmath}
\usepackage{latexsym}
\usepackage{amsthm, amscd, amsfonts, amssymb}
\usepackage{multirow}
\usepackage{makecell}
\usepackage{dcolumn}
\usepackage{tabularx}
\usepackage{xcolor}
\definecolor{RED}{RGB}{255,0,0}
\definecolor{BLUE}{RGB}{0,0,255}

\usepackage{etoolbox}
\usepackage{threeparttable}

\begin{document}

\title{{Higher-Order Modulational Instability in Spin-Orbit-Coupled
Bose-Einstein Condensates: {Finite-Raman Effects and Branch-Resolved Spectral
Redistribution}}}
\author{Cui-Cui Ding$^1$}
\author{Boris A. Malomed$^{2,3}$}
\author{Qin Zhou$^{1,4}$}
\email{qinzhou@whu.edu.cn}
\affiliation{$^1$
School of Microelectronics, Wuhan Textile University, Wuhan
430200, China}

\affiliation{$^2$Department of Physical Electronics, School of Electrical Engineering,
	Faculty of Engineering, Tel Aviv University, Tel Aviv, Israel}
\affiliation{$^3$Instituto de Alta Investigaci\'{o}n, Universidad de Tarapac\'{a},
Casilla 7D, Arica, Chile}
\affiliation{$^4$State Key Laboratory of New Textile Materials
and Advanced Processing, Wuhan Textile University, Wuhan 430200, China}


\begin{abstract}
We investigate higher-order modulational instability and the nonlinear
spectral redistribution in a spin-orbit-coupled (SOC) binary Bose--Einstein
condensate with equal interactions. {For finite Raman coupling, the
system is nonintegrable and exhibits both nonoscillatory and oscillatory
instability regimes.} Raman coupling reshapes the instability bands and
changes what harmonics can grow. Direct simulations show that this
reorganization modifies the subsequent nonlinear sideband dynamics.
In the zero-Raman limit, SOC shifts successive harmonics
between instability bands and stability gaps at a fixed carrier wavenumber.
It thereby controls higher-order instability, distinct breather branches,
and frequency-hopping sequences. {We derive exact algebraic coefficients for
the carrier and arbitrary-order sidebands.} {This yields an exact geometric
power spectrum for the complete infinite sideband sequence and a unified
description of carrier depletion, directional imbalance, and sideband decay,
including an exact relation between the instability gain and the minimum
spectral decay rate.}
For two coexisting breather branches with the same background and
modulation wavenumber, {the higher-gain branch can have a smaller maximum
carrier depletion, a slower high-order spectral decay, and a weaker directional
imbalance.} {SOC can also restore a symmetric sideband
spectrum when the SOC-induced momentum shift compensates the carrier
wavenumber, even when both $k$ and $k_L$ are nonzero.} These analytical and numerical
results clarify how SOC and Raman couplings organize higher-order
instability and nonlinear spectral redistribution, providing a unified
framework for studying coherent nonlinear-wave dynamics in multicomponent
condensates.\end{abstract}

\maketitle

\section{INTRODUCTION}

Modulational instability (MI) amplifies weak perturbations on a
plane-wave background in optics, plasmas, fluids, and
Bose--Einstein condensates
(BECs)~\cite{Zakharov2009,Segur2007,Everitt2017,Otajonov2022}.
Dispersion management, nonlocal interactions, and dissipation can
modify MI, while optical and fluid systems also exhibit more complex
instability regimes~\cite{Armaroli2023,Tabi2022,Kharif2010,Perego2022,Vanderhaegen2021}.
Beyond linear growth, nonlinear coupling transfers power from the
carrier to sidebands and can generate localized
structures~\cite{Kraych2019,Chin2015}.
Akhmediev breathers (ABs), which are periodic in space and localized
in time, provide exact solutions describing coherent MI growth and
decay~\cite{Chen2021,Pierangeli2018}.
The associated power exchange is closely related to
Fermi--Pasta--Ulam--Tsingou (FPUT) recurrence, in which sideband
power returns to the
carrier~\cite{Kimmoun2016,Pierangeli2018,Yin2023,Chowdury2025}.
Higher-order MI involves higher harmonics of the initial modulation
that also lie within the instability bands and participate in the
nonlinear evolution.
Their nonlinear interaction can produce multiple breather events
and spectral cascades~\cite{Erkintalo2011,Chen2023}.

Intercomponent coupling allows a wider range of MI dynamics in
vector systems~\cite{Liu2023a,Madimabe2023,Mossman2025}.
{The Manakov system supports asymmetric AB spectra and nondegenerate
breather branches, while stable gaps between instability bands can
lead to frequency hopping}~\cite{Chen2021,Chen2023}.
Spin-orbit coupling (SOC) couples atomic pseudospin to momentum
and provides tunable single-particle dispersion in
BECs~\cite{Lin2011,SOC-Review1,SOC-Review2,SOC-Review3}.
Related studies have examined solitons and breathers in helicoidal
SOC-BECs~\cite{Ding2024,Ding2025,Ding2026}, as well as periodic
modulation of breather phases and backgrounds in coupled
condensates~\cite{fei2026}.
{For finite Raman coupling, Bhat et al. analyzed MI with general
interaction coefficients}~\cite{Bhat2015}, { while Mithun and Kasamatsu
studied its nonlinear dynamics}~\cite{Mithun2019}.
MI has also been investigated in condensates with helicoidal SOC,
residual interactions, and open-system
dynamics~\cite{Tabi2023,Tabi2022a,Otlaadisa2021}.

{Here we examine how SOC affects higher-order MI and nonlinear
spectral redistribution in a binary BEC with equal interactions.}
{For finite Raman coupling, changing the SOC strength modifies the
instability spectrum and determines which harmonics of a specified
modulation are unstable.}
In the integrable zero-Raman limit, {varying SOC at a fixed carrier
wavenumber shifts successive harmonics between instability bands
and stable gaps and changes the available breather branches.}
{Direct simulations and exact multi-AB solutions show how these
SOC-induced changes affect the redistribution of sideband power,
nondegenerate breather excitation, and frequency hopping.}

{In the zero-Raman limit, the equations can be mapped to the
Manakov system by the gauge transformation.}
{The known exact Fourier spectrum of elementary vector ABs requires
derivatives of increasing order for positive
sidebands}~\cite{Chen2021}, complicating direct analysis of the full
sideband sequence.
{We overcome this limitation by deriving exact algebraic coefficients
without order-dependent differentiation.}
{These expressions establish exact spectral laws for the entire
infinite sequence of sidebands, providing a unified analytical
description of carrier depletion, directional imbalance, and
spectral decay.}
Specifically, we derive exact expressions for the common geometric
decay rate of positive and negative sidebands and their
order-independent asymmetry ratio, and establish an exact relation
between the MI growth rate and the minimum spectral decay rate.
We further obtain analytical conditions for spectral symmetry,
one-sided spectra, and complete carrier depletion.
{A comparison of two coexisting AB solutions shows that the higher-gain
branch can undergo less maximum carrier depletion, while its high-order
sidebands decay more slowly and show weaker directional imbalance.}

{The paper is organized as follows.
Section~II introduces the SOC-BEC model, analyzes MI with and without
Raman coupling, and presents the higher-order MI phase diagrams.
Section~III compares direct simulations with exact multi-AB solutions
in the zero-Raman limit and examines nonlinear dynamics at finite
Raman coupling.
Section~IV derives the exact Fourier spectrum and analyzes carrier
depletion, directional asymmetry, and spectral decay, including their
relations to the MI gain.
Section~V summarizes the main results.}

\section{THE MODEL, MODULATIONAL INSTABILITY, AND EXACT MULTI-BREATHER
SOLUTIONS}

\subsection{The Spin-Orbit-Coupled BEC System}

We consider a quasi-one-dimensional binary BEC with equal-weight
Rashba--Dresselhaus SOC~\cite{Bychkov1984,Dresselhaus1955}.
Tight transverse harmonic confinement with frequency $\omega_{\bot}$
freezes the transverse degrees of freedom, allowing an effective
one-dimensional mean-field description.
The energy density is~\cite{Achilleos2013}
\begin{equation}
\begin{aligned}
\mathcal{E}_0={}&
\mathbf{\Psi}^{\dagger}\mathcal{H}_s\mathbf{\Psi}
+\frac{g_{11}}{2}|\Psi_1|^4
+\frac{g_{22}}{2}|\Psi_2|^4\\
&+g_{12}|\Psi_1|^2|\Psi_2|^2,
\end{aligned}
\label{energy_functional}
\end{equation}
where $\mathbf{\Psi}=(\Psi_1,\Psi_2)^T$ is the spinor order parameter.
The single-particle Hamiltonian is
\begin{equation}
\mathcal{H}_s=
\frac{1}{2m}(p_x\mathbf{I}-k_L\boldsymbol{\sigma}_z)^2
+\Omega_R\boldsymbol{\sigma}_x+V(x)\mathbf{I},
\label{single_particle_hamiltonian}
\end{equation}
where $p_x=-i\hbar\partial_x$ is the longitudinal momentum operator,
$m$ is the atomic mass, $\boldsymbol{\sigma}_{x,z}$ are the Pauli
matrices, and $\mathbf{I}$ is the identity matrix.
Here $k_L$ characterizes the SOC strength, $\Omega_R$ is the Raman
coupling strength between the two pseudospin components, and $V(x)$
is the longitudinal external potential.
Throughout this work, we consider $V(x)=0$.
The effective one-dimensional interaction coefficients are
$g_{ij}=2\alpha_{ij}\hbar\omega_{\bot}$, where $\alpha_{ij}$ are the
corresponding $s$-wave scattering lengths.
We assume equal interactions, $g_{11}=g_{22}=g_{12}=g$.

We measure length, time, energy, and one-dimensional density in units
of $a_{\bot}=\sqrt{\hbar/(m\omega_{\bot})}$,
$\omega_{\bot}^{-1}$, $\hbar\omega_{\bot}$, and
$(2|\alpha_{11}|)^{-1}$, respectively.
The coupling parameters are rescaled as
$k_L\mapsto-a_{\bot}k_L/\hbar$ and
$\Omega_R\mapsto\Omega_R/(\hbar\omega_{\bot})$,
with the same symbols used for the dimensionless quantities.
With these scalings and the constant SOC energy shift absorbed
into the energy zero, the coupled Gross--Pitaevskii equations
take the form~\cite{Achilleos2013}
\begin{equation}
\begin{aligned}
i\frac{\partial\Psi_1}{\partial t}
={}&-\frac{1}{2}\frac{\partial^2\Psi_1}{\partial x^2}
-ik_L\frac{\partial\Psi_1}{\partial x}\\
&-\sigma(|\Psi_1|^2+|\Psi_2|^2)\Psi_1+\Omega_R\Psi_2,\\
i\frac{\partial\Psi_2}{\partial t}
={}&-\frac{1}{2}\frac{\partial^2\Psi_2}{\partial x^2}
+ik_L\frac{\partial\Psi_2}{\partial x}\\
&-\sigma(|\Psi_1|^2+|\Psi_2|^2)\Psi_2+\Omega_R\Psi_1,
\end{aligned}
\label{model_equation}
\end{equation}
where $\sigma=-\operatorname{sgn}(\alpha_{11})$, with
$\sigma=+1$ and $-1$ corresponding to attractive and repulsive
interactions, respectively.
The zero-Raman limit is integrable and admits the exact elementary
and multi-AB solutions used below.

\subsection{{Modulational Instability with Finite Raman Coupling}}

Building on the MI analysis in Ref.~\cite{Bhat2015},
we investigate
how SOC and finite Raman coupling determine the distribution
of unstable harmonics and construct the corresponding
higher-order MI phase diagrams.
These diagrams identify the instability regimes underlying
the nonlinear evolution and breather excitation studied below.
In the zero-Raman limit, we further connect higher-order MI
with exact multi-AB solutions and derive the exact Fourier
spectra of elementary ABs.
For the finite-Raman analysis, we consider an equal-density,
in-phase background with zero carrier wave number,
\begin{equation}
\Psi_{10}=\Psi_{20}=\sqrt{N}\,e^{i\mu t},
\qquad \mu=2\sigma N-\Omega_R,
\label{finite_Raman_background}
\end{equation}
where $N$ is the density of each component.

To examine the linear stability of this background, we introduce
small perturbations,
\begin{equation}
\begin{aligned}
\Psi_j=\Psi_{j0}\big[&1+p_j e^{-i(\omega x-\varOmega t)}\\
&+\chi_j^{*}e^{i(\omega x-\varOmega^{*}t)}\big],
\qquad j=1,2,
\end{aligned}
\label{finite_Raman_perturbation}
\end{equation}
where $\omega$ is the modulation wavenumber, $\varOmega$ is the
temporal eigenfrequency, and $p_j$ and $\chi_j$ are small complex
amplitudes.
{The initial perturbations in the two components are not assumed
to be identical: the component-dependent coefficients $p_j$
and $\chi_j$ allow different amplitudes and phases.}
{Both components belong to the same normal mode and share
$\omega$ and $\varOmega$, while their relative perturbation
amplitudes and phases are determined by the corresponding
eigenvector.}

Substituting Eq.~(\ref{finite_Raman_perturbation}) into
Eq.~(\ref{model_equation}) and keeping only linear terms yields
four homogeneous equations for the perturbation amplitudes.
Requiring nontrivial solutions gives the dispersion
relation~\cite{Bhat2015}
\begin{equation}
\varOmega^4-T\varOmega^2+D=0,
\label{finite_Raman_dispersion}
\end{equation}
where
{\begin{equation}
\begin{aligned}
T={}&\frac{\omega^2}{2}
\left(\frac{\omega^2}{2}-4\sigma N\right)
+\left(\frac{\omega^2}{2}-2\Omega_R\right)^2
+2k_L^2\omega^2,\\
D={}&D_1D_2=
\left[
\frac{\omega^2}{2}
\left(\frac{\omega^2}{2}-2\Omega_R\right)
-k_L^2\omega^2
\right]\\
&\times\left[
\left(\frac{\omega^2}{2}-4\sigma N\right)
\left(\frac{\omega^2}{2}-2\Omega_R\right)
-k_L^2\omega^2
\right].
\end{aligned}
\label{finite_Raman_TD}
\end{equation}}

The four eigenfrequency branches are
{\begin{equation}
\varOmega_{1,2,3,4}=\pm\sqrt{\frac{T\pm\sqrt{\Delta_R}}{2}}.
\label{finite_Raman_roots}
\end{equation}}
where $\Delta_R=T^2-4D$.
MI occurs when at least one branch has a nonzero imaginary part.
All branches are real when
\begin{equation*}
\Delta_R\geq0,\qquad T\geq0,\qquad D\geq0.
\end{equation*}
The maximum MI growth rate,
$G_{\max}=\max_{j=1,\ldots,4}|\operatorname{Im}\varOmega_j|$,
is
{\begin{equation}
G_{\max}=
\begin{cases}
\displaystyle
\sqrt{\max\!\left(0,\frac{\sqrt{\Delta_R}-T}{2}\right)},
& \Delta_R\geq0,\\[2ex]
\displaystyle
\frac{1}{2}\sqrt{2\sqrt{D}-T},
& \Delta_R<0,
\end{cases}
\label{finite_Raman_gain}
\end{equation}}
where $\Delta_R<0$ implies $D>0$.

\begin{figure}[th]
\includegraphics[width=0.485\columnwidth]{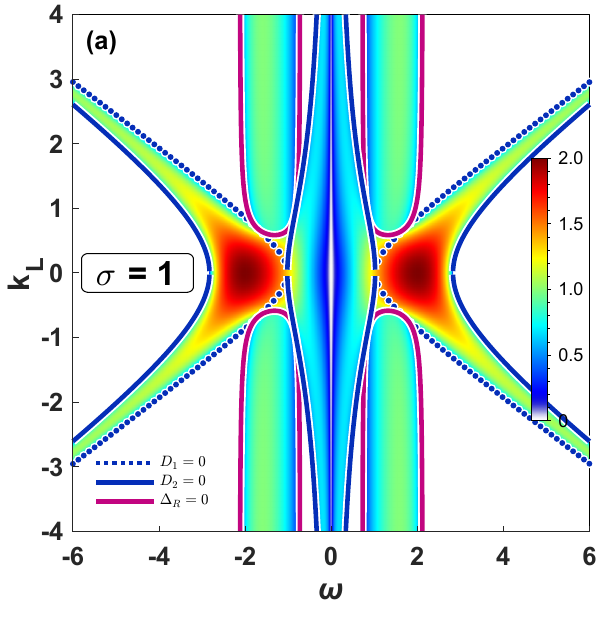}\hfill
\includegraphics[width=0.485\columnwidth]{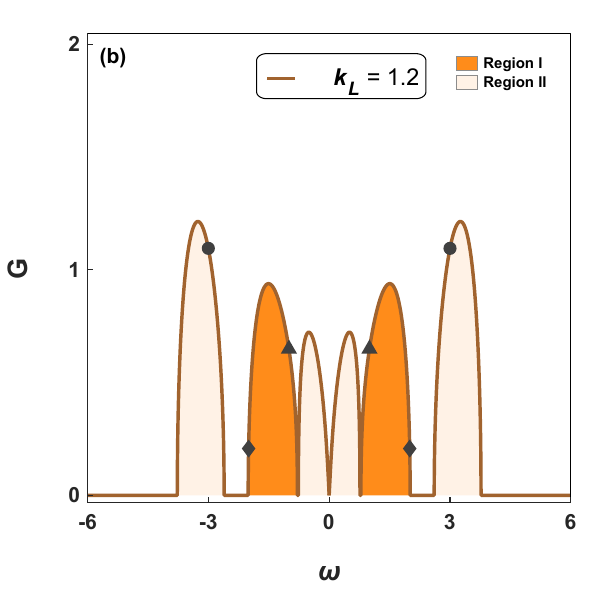}\newline
\includegraphics[width=0.98\columnwidth]{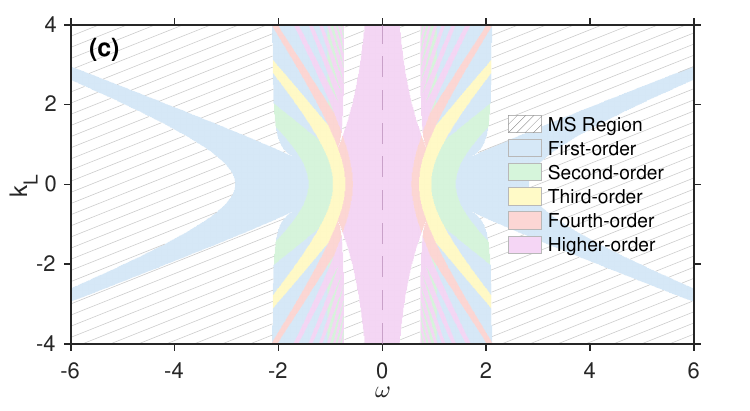}\newline
\vspace{-3mm}
\caption{Finite-Raman MI structure for $N=\sigma=1$,
$\Omega_R=0.26$, and $k=0$. (a) Maximum MI gain $G_{\max}$ in the
$(\omega,k_L)$ plane. The curves mark $D_1=0$, $D_2=0$, and $\Delta_R=0$.
(b) Gain profile at $k_L=1.2$; the shaded regions indicate the corresponding
unstable eigenvalue structures, and the symbols mark the reference
fundamental $\omega_0=1$ and its harmonics. (c) Highest unstable harmonic
index $n$ in the $(k_L,\omega)$ plane; hatching denotes a stable fundamental.}
\label{fig:finite_Raman_MI}
\end{figure}

{Figure~\ref{fig:finite_Raman_MI}(a) should be interpreted together
with the sign conditions in Eq.~(\ref{finite_Raman_roots}). The white
regions have $G_{\max}=0$, with all four eigenfrequencies being real, and
the stable plane wave, whereas the colored regions have
$G_{\max}>0$, exhibiting MI. The curves $D_1=0$, $D_2=0$, and
$\Delta_R=0$ mark changes in the algebraic structure of the
eigenfrequency branches. They do not all represent stability
boundaries, because neighboring regions can both be unstable while
having different eigenfrequency structures. At $\Delta_R<0$, the
four eigenfrequencies form a complex quartet, corresponding to
oscillatory MI. When $\Delta_R>0$ and $D<0$, one pair is real and the
other is purely imaginary,corresponding to exponentially growing (non-oscillatory) MI. Further,
an unstable structure occurs too at $\Delta_R>0$, $D>0$, and $T<0$, where
both pairs are purely imaginary. This structure is present in the
full map, although it is not crossed by the $k_L=1.2$ cut shown in
Fig.~\ref{fig:finite_Raman_MI}(b).

Figure~\ref{fig:finite_Raman_MI}(a) shows the MI gain map for
$N=\sigma=1$ and $\Omega_R=0.26$. The cut at $k_L=1.2$ in
Fig.~\ref{fig:finite_Raman_MI}(b) directly illustrates the different
instability types: the fundamental frequency $\omega_0=1$ and its second
harmonic lie in the complex-quartet region (Region~I), the third
harmonic lies in the single-imaginary-pair region (Region~II), while
the fourth harmonic is stable. Thus, different harmonics generated
by the same fundamental modulation can experience different
eigenfrequency structures and different types of MI. The resulting
higher-order MI phase diagram is shown in
Fig.~\ref{fig:finite_Raman_MI}(c) and analyzed in Sec.~II D.}

\subsection{{Integrable Zero-Raman Limit and Exact Multi-AB Solutions}}

{In the zero-Raman limit, Eq.~(\ref{model_equation}) can be mapped
to the Manakov system by the gauge transformation.} The first-order derivative
terms are analogous to walk-off terms in coupled optical-fiber
models~\cite{Sun2024} and can be removed by a component-dependent
gauge transformation. {The inverse transformation restores the
SOC-dependent phases and effective wave numbers.}
The elementary AB solution takes the form
\begin{equation}
\Psi_j=\Psi_{j0}
\frac{\cosh(\Theta+i\vartheta_j)e^{i\theta_{1j}}
+\varpi\cos(\Upsilon-i\upsilon_j)e^{i\theta_{2j}}}
{\cosh\Theta+\varpi\cos\Upsilon}.
\label{breather}
\end{equation}
Here $\Psi_{j0}$ is the equal-density counterpropagating
background,
{\begin{equation}
\begin{aligned}
\Psi_{j0}&=\sqrt{N}\exp[i(k_jx+\mu t)],
\qquad k_1=-k_2=k,\\
\mu&=2\sigma N-\frac{k^2}{2}-k_Lk.
\end{aligned}
\label{plane_wave_solution}
\end{equation}}
The effective background wave numbers are
$b_1=-b_2=k+k_L$.
For matched transformed initial and boundary data, the
modulation dynamics depends on $k$ and $k_L$ only through
their sum.
{At a fixed carrier wavenumber $k$, varying the SOC strength
changes the MI bands and available AB branches, allowing
harmonics of a specified modulation to enter or leave
unstable regions.}

The temporal localization and spatial periodicity are
described by
\begin{equation}
\begin{aligned}
\Theta&=\omega\tau_I t_1,\\
\Upsilon&=\omega\left[x_1+
\left(\tau_R+\frac{\omega}{2}\right)t_1\right]
+\arg\frac{2\tau_I}{2\tau_I-i\omega},
\end{aligned}
\label{parameters}
\end{equation}
where $\omega$ is the modulation wavenumber,
$x_1=x-x_{10}$, $t_1=t-t_{10}$, and $\tau_R,\tau_I$
are the real and imaginary parts of $\tau$.
The eigenvalue $\tau$ satisfies
\begin{equation}
1+\sigma\sum_{j=1}^{2}
\frac{N}{(\tau+b_j)(\tau+\omega+b_j)}=0.
\label{characteristic_equation}
\end{equation}
The remaining parameters and the multi-AB construction
are given in Appendix A.
These solutions describe selected coherent MI events;
broadband perturbations can lead to more complex
evolution~{\cite{Suret2024}}.

For the background in Eq.~(\ref{plane_wave_solution}),
the zero-Raman MI eigenfrequencies are
\begin{equation}
\varOmega_{1,2,3,4}=\pm |\omega |\sqrt{\Lambda \pm \sqrt{\Delta }},
\label{Omega}
\end{equation}
where $b=k+k_L$ and
\begin{equation}
\Lambda=b^2-\sigma N+\frac{\omega^2}{4},
\qquad
\Delta=-4\sigma Nb^2+N^2+b^2\omega^2.
\label{zero_Raman_Lambda_Delta}
\end{equation}
The corresponding roots of
Eq.~(\ref{characteristic_equation}) are
\begin{equation}
\tau_{1,2,3,4}
=\pm\sqrt{\Lambda\pm\sqrt{\Delta}}-\frac{\omega}{2}.
\label{tau}
\end{equation}
For regular AB solutions, only nonreal roots satisfying
Eq.~(\ref{characteristic_equation}) with
$(\tau+b_j)(\tau+b_j+\omega)\ne0$ for both components
are admissible.

{For $\omega>0$, the MI eigenfrequency and the corresponding
AB eigenvalue satisfy
$\varOmega=\omega(\tau+\omega/2)$.
Since $\omega/2$ is real, the MI growth rate is
$|\operatorname{Im}\varOmega|=|\omega\tau_I|$.}
This rate also governs the exponential growth and decay
of the breather modulation near the plane-wave background.

\subsection{{Higher-Order MI Channels and Phase Diagrams}}

{For a given fundamental wavenumber $\omega>0$,
we define the highest unstable harmonic index as}
\begin{equation*}
n=\max\left\{m\in\mathbb{N}:G_{\max}(m\omega)>0\right\},
\qquad G_{\max}(\omega)>0,
\end{equation*}
{where $G_{\max}(m\omega)$ is the largest MI growth rate
among the four branches belonging to the $m$-th harmonic.
Zero-gain band edges are excluded, and $n$ is undefined
when the fundamental is stable.
Intermediate harmonics may lie in stable gaps, so $n$
does not need to be equal to the number of unstable harmonics.
It also differs from the multi-AB order, which counts
breather constituents and can exceed $n$ when a harmonic
supports multiple unstable branches.}
This definition also applies to the finite-Raman phase
diagram in Fig.~\ref{fig:finite_Raman_MI}(c).

We now examine the zero-Raman MI phase diagrams in the
$(b,\omega)$ plane for repulsive and attractive interactions.
We fix the carrier wave number at $k=0.2$ and vary the
SOC strength $k_L$, which shifts the effective background
wave number and changes what harmonics are unstable.

\begin{figure}[th]
\includegraphics[width=0.485\columnwidth]{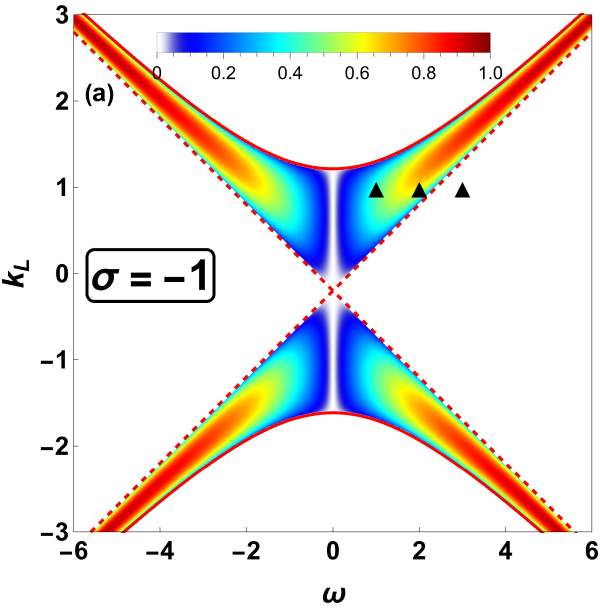}\hfill
\includegraphics[width=0.485\columnwidth]{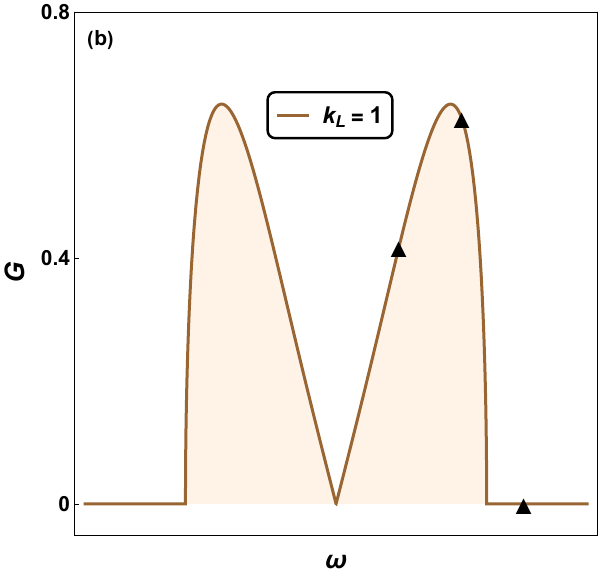}\newline
\includegraphics[width=0.98\columnwidth]{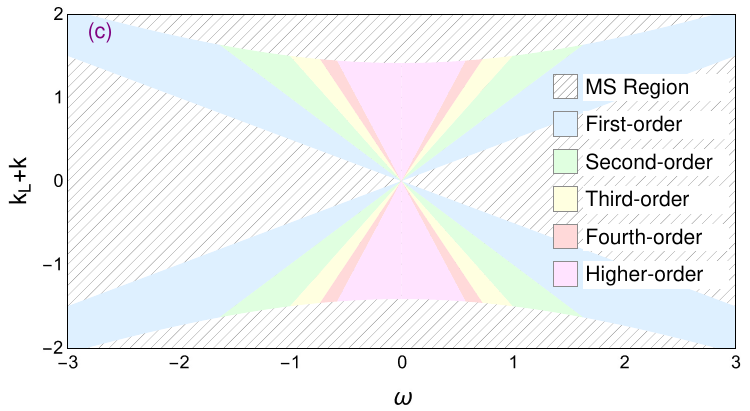}\newline
\vspace{-3mm}
\caption{{Repulsive interaction: (a) MI gain distribution; (b) gain versus
modulation wavenumber for representative SOC parameters, with triangles
marking $\protect\omega=1$ and its harmonics; (c) highest unstable harmonic
index $n$ in the $(k_L+k,\protect\omega)$ plane, with hatching denoting a
stable fundamental. Other parameters are $N=1$ and $k=0.2$.}}
\label{fig:repulsive_MI}
\end{figure}

For repulsive interactions ($\sigma=-1$), only the branches
with the inner minus sign in Eq.~(\ref{Omega}) can become
unstable.
The MI gain is symmetric about $\omega=0$ and $k_L=-k$
[Fig.~\ref{fig:repulsive_MI}(a)].
Evaluating the dispersion relation at $m\omega$, the
$m$th harmonic is unstable when
{\begin{equation*}
2\sqrt{\max(b^2-2N,0)}<m\omega<2|b|.
\end{equation*}}
{When the fundamental is unstable, the unstable harmonics
form a consecutive sequence, with highest index}
{\begin{equation*}
n=\left\lceil\frac{2|b|}{\omega}\right\rceil-1.
\end{equation*}}
The strict inequalities exclude harmonics at the neutral
band edges.

For $N=1$ and $k_L=1$ in Fig.~\ref{fig:repulsive_MI}(b), the instability
condition becomes $0<m\omega<2.4$.
With $\omega=1$, the first two harmonics are unstable
and all higher harmonics are stable, giving $n=2$.
At fixed carrier wavenumber, varying the SOC strength
moves the harmonics across the instability boundaries,
producing the stepwise changes in $n$ shown in
Fig.~\ref{fig:repulsive_MI}(c).

\begin{figure}[th]
\includegraphics[width=0.485\columnwidth]{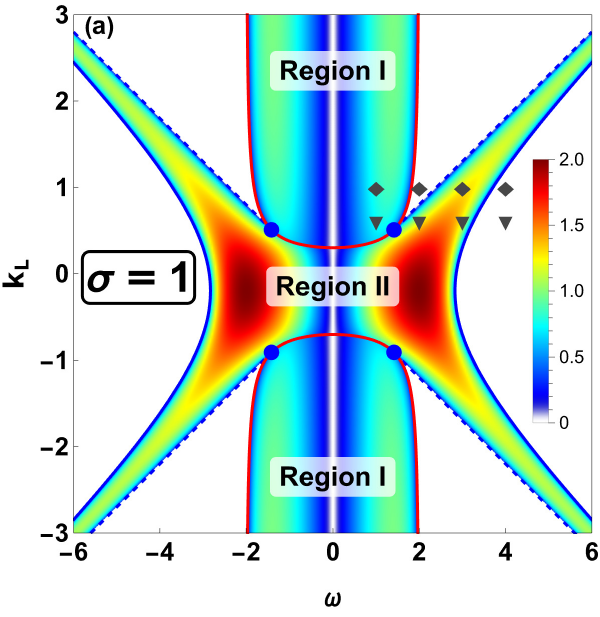}\hfill
\includegraphics[width=0.485\columnwidth]{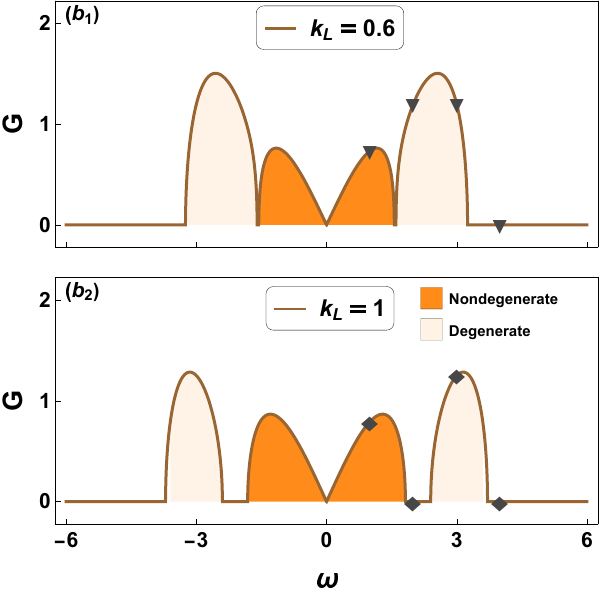}\newline
\includegraphics[width=0.98\columnwidth]{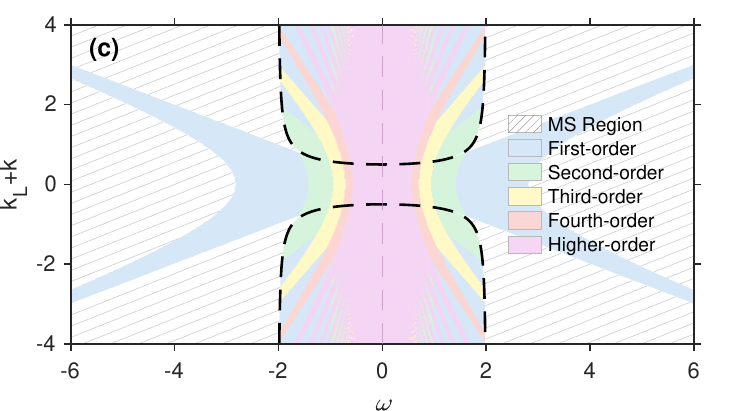}\newline
\vspace{-3mm}
\caption{{Attractive interaction: (a) MI gain distribution; (b) gain versus
modulation wavenumber for representative SOC parameters, with triangles and
rhombuses marking $\protect\omega=1$ and its harmonics; (c) highest unstable
harmonic index $n$ in the $(k_L+k,\protect\omega)$ plane. Regions I and II
denote the drifting nondegenerate and stationary-modulation branch regimes,
respectively; hatching denotes a stable fundamental and black dashed curves
mark $\Delta=0$. Other parameters are $N=1$ and $k=0.2$.}}
\label{fig:attractive_MI}
\end{figure}

{For attractive interactions ($\sigma=+1$),
Fig.~\ref{fig:attractive_MI}(a) contains two regimes with unstable branches.}
For an AB associated with the $m$th harmonic, the phase
velocity of its periodic modulation, obtained from
$\Upsilon=\mathrm{const}$, is $-(\tau_R+m\omega/2)$.
Here $\tau$, $\Lambda$, and $\Delta$ are evaluated at $m\omega$.
In Region I ($\Delta<0$), two nondegenerate AB branches
have equal MI growth rates and opposite modulation
phase velocities~{\cite{chong2022}}.
In the unstable part of Region II ($\Delta>0$),
$\tau_R=-m\omega/2$, so the modulation phase velocity
vanishes.
One or two unstable branches can occur; when both
$\Lambda\pm\sqrt{\Delta}$ are negative, two branches
coexist with different growth rates.

For $b^2<N/2$, the $m$th harmonic is unstable when
{\begin{equation*}
0<m\omega<2\sqrt{b^2+2N}.
\end{equation*}}
For $b^2\geq N/2$, the condition is
{\begin{equation*}
0<m\omega<\sqrt{4N-\frac{N^2}{b^2}}
\quad\text{or}\quad
2|b|<m\omega<2\sqrt{b^2+2N}.
\end{equation*}}
At $b^2=N/2$, only the neutral point
$m\omega=\sqrt{2N}$ separates the two intervals;
a finite stability gap opens for $b^2>N/2$.
{Thus, varying SOC at fixed carrier wavenumber can
split the MI band into two separated intervals.}

Within the unstable region, $\Delta=0$ separates AB
branches with moving and stationary modulation phases.
This condition differs from $b^2>N/2$, which determines
whether a finite stability gap exists.

Fig.~\ref{fig:attractive_MI}(c) follows by applying the instability
conditions to each harmonic $m\omega$.
{In the attractive case, intermediate harmonics can lie
in the stable gap while higher harmonics remain unstable.
These stable harmonics do not grow independently through
linear MI, but can acquire spectral power through
nonlinear mixing.}
The separation of unstable harmonics provides the linear
mechanism underlying the frequency-hopping example in Sec.~III.
The black dashed curves mark $\Delta=0$ evaluated at
the fundamental wavenumber.

\begin{figure}[th]
\centering
\includegraphics[width=0.45\textwidth]{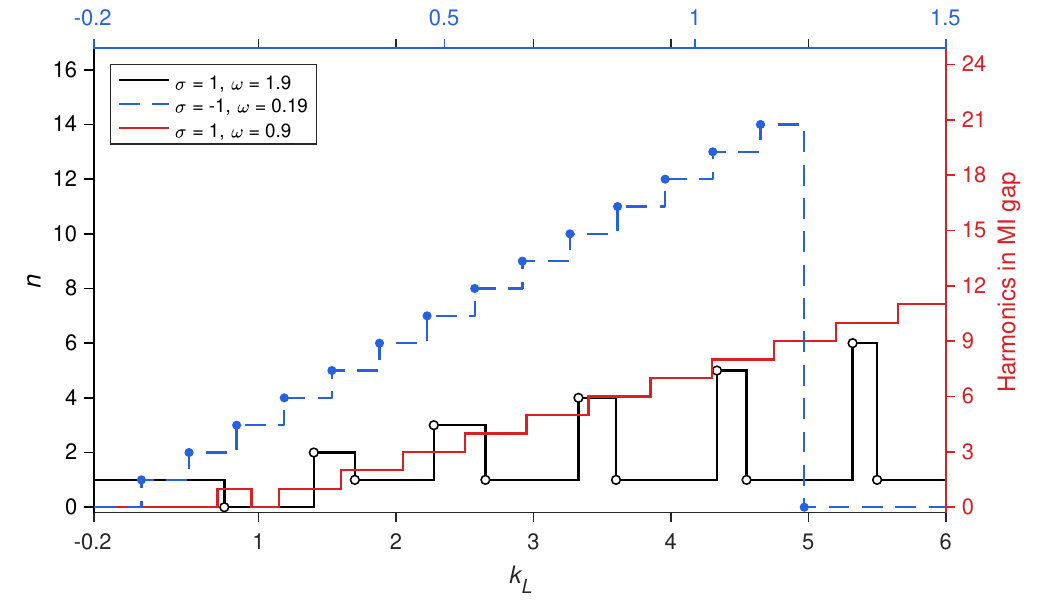}\vspace{-2mm}
\caption{{Highest unstable harmonic index $n$ (left axis) and number of
harmonics lying in the stability gap (right axis) versus $k_L$. Black solid: attractive,
$\protect\omega=1.9$; blue dashed: repulsive, $\protect\omega=0.19$; red
solid: gap-harmonic count for attractive interaction,
$\protect\omega=0.9$. A zero on an $n$ curve is used only as a plotting
marker for a stable fundamental; no value of $n$ is assigned there. The
black/red curves use the lower $k_L$ axis and the blue curve the upper axis.
Other parameters are $N=1$ and $k=0.2$.}}
\label{fig:harmonic_order}
\end{figure}

Fig.~\ref{fig:harmonic_order} shows how the harmonic distribution
changes with SOC strength.
For repulsive interactions, the unstable harmonics are
consecutive and $n$ changes in steps.
{For attractive interactions, the stable gap makes the
unstable harmonics nonconsecutive, and $n$ can vary
nonmonotonically.}
For $b^2>N/2$, the red curve counts harmonics inside
the stable gap,
\begin{equation*}
\sqrt{4N-\frac{N^2}{b^2}}<m\omega<2|b|.
\end{equation*}
The strict inequalities exclude the neutral band edges.

These linear results identify which harmonics can grow
through MI, but do not determine their subsequent
nonlinear evolution.
Section~III examines this evolution through direct
simulations and selected exact multi-AB solutions.

\section{NONLINEAR EVOLUTION AND MULTI-BREATHER DYNAMICS}

\subsection{{Numerical Validation of Selected Multi-Breather Dynamics}}

{In the zero-Raman limit, we compare direct simulations
of Eq.~(\ref{model_equation}) with selected exact
multi-AB solutions.}
The comparison examines the number of breather events,
their modulation wavenumbers, localization sequence,
and spectral redistribution.
The multi-AB order $M$ counts the breather constituents
and need not equal the highest unstable harmonic index
$n$, since one harmonic can support multiple unstable
branches.

The simulations start from the same weak relative
modulation in both components,
\begin{equation}
\Psi_j(x,0)=[1+\epsilon\cos(\omega x)]\Psi_{j0}(x,0),
\label{initial_condition}
\end{equation}
with $\epsilon=0.002$, and use a split-step Fourier method.
The initial modulation seeds the fundamental, while
higher harmonics are generated by nonlinear evolution
and may lie in stable or unstable MI regions.

\begin{figure*}[tbp]
\centering
\includegraphics[width=0.7\textwidth]{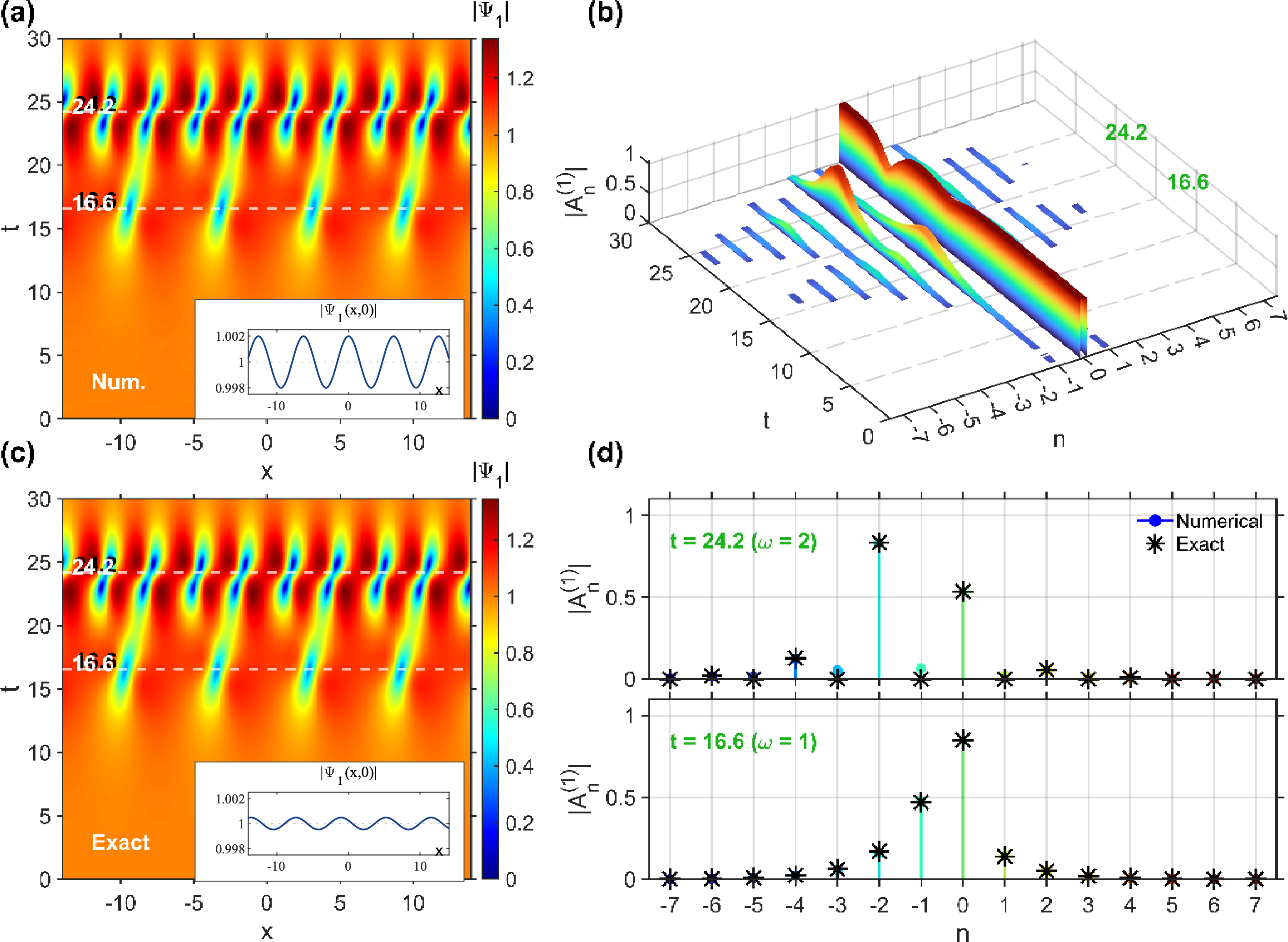}\vspace{-2mm}
\caption{{Repulsive interaction ($\protect\sigma=-1$) for $k_L=1$,
with highest unstable harmonic index $n=2$. (a,c) Numerical and exact evolution
of} {$|\Psi _1|$} for the selected $M=2$ multi-AB
excitation $\{\protect\omega,2\protect\omega\}$.
The insets in (a) and (c) show $|\Psi _1(x,0)|$ with identical
axis limits and a vertically magnified view around the background
amplitude. {(b) Spatiotemporal Fourier
spectrum showing a cascade through successive sidebands. (d) Numerical spectra
and the corresponding elementary-AB spectral benchmarks near the constituent
localization times.}}
\label{fig:repulsive_multiAB}
\end{figure*}

\begin{figure*}[tbp]
\centering
\includegraphics[width=0.7\textwidth]{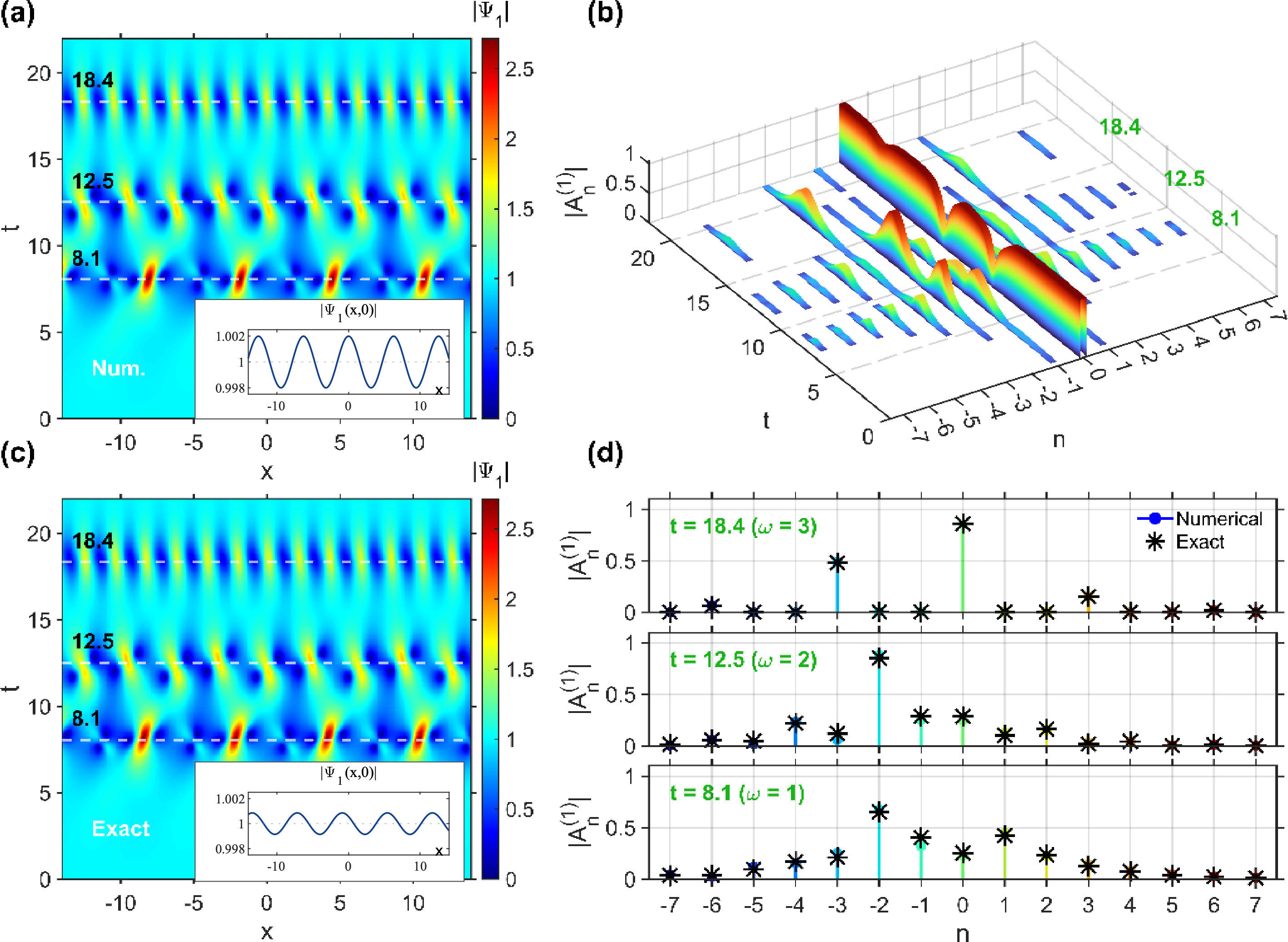}\vspace{-2mm}
\caption{{Attractive interaction ($\protect\sigma=1$) for $k_L=0.6$,
with highest unstable harmonic index $n=3$. (a,c) Numerical and exact evolution
of} {$|\Psi _1|$} for the selected $M=4$ multi-AB
excitation $\{\protect\omega,\protect\omega,2\protect\omega,3\protect\omega\}$.
The insets in (a) and (c) show $|\Psi _1(x,0)|$ with identical
axis limits and a vertically magnified view around the background
amplitude. {(b) Spatiotemporal Fourier
spectrum showing the sideband cascade. (d) Numerical spectra and the
corresponding Fourier projections of the selected multi-AB solution near the
constituent localization times.}}
\label{fig:attractive_multiAB}
\end{figure*}

\begin{figure*}[tbp]
\centering
\includegraphics[width=0.7\textwidth]{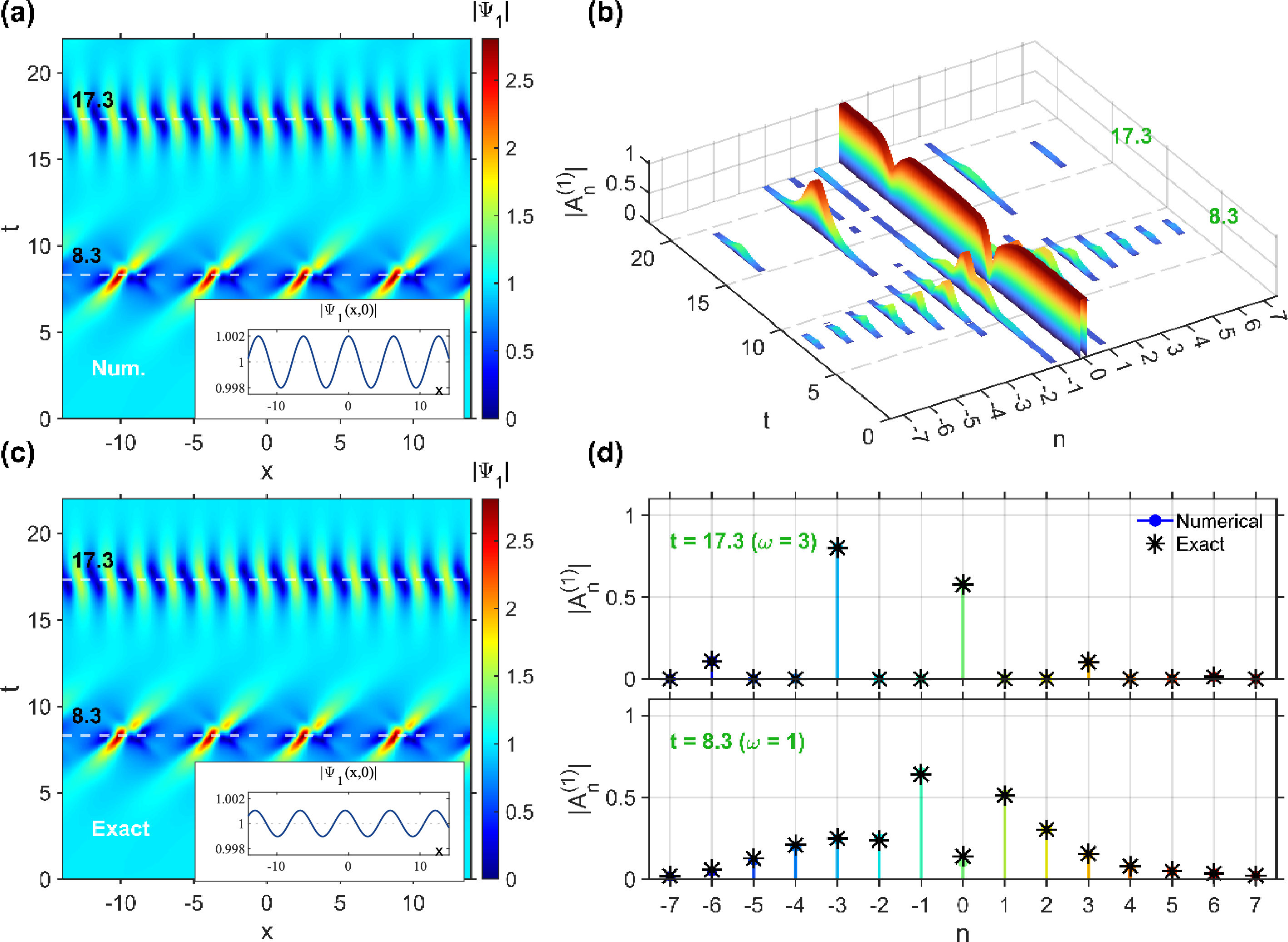}\vspace{-2mm}
\caption{{Frequency hopping for the attractive interaction
($\protect\sigma=1$) at $k_L=1$, with highest unstable harmonic index $n=3$.
(a,c) Numerical and exact evolution of} {$|\Psi _1|$}
for the selected $M=3$ multi-AB excitation
$\{\protect\omega,\protect\omega,3\protect\omega\}$.
The insets in (a) and (c) show $|\Psi _1(x,0)|$ with identical
axis limits and a vertically magnified view around the background amplitude.
{The $2\protect\omega$ harmonic in the stability gap is not an independent MI breather channel,
although nonlinear mixing can leave a nonzero Fourier component at this
wavenumber.} {(b) Spatiotemporal Fourier
spectrum showing the corresponding sideband redistribution. (d) Numerical
spectra and the corresponding Fourier projections of the selected multi-AB
solution near the constituent localization times.}}
\label{fig:frequency_hopping}
\end{figure*}

The numerical initial condition differs from the exact
multi-AB profiles, as shown by the insets in
Figs.~\ref{fig:repulsive_multiAB}--\ref{fig:frequency_hopping} using identical axis limits.
For each comparison, the constituent wavenumbers and
branches are selected from the MI analysis, while
spatial and temporal shifts locate the individual
breather events.
The comparison therefore concerns selected coherent
growth--decay events.
It does not establish a general description of
long-time MI dynamics under broadband perturbations,
which can involve integrable turbulence or soliton
gases~{\cite{Suret2024}}.

The equal-density counterpropagating background and the
even modulation in Eq.~(\ref{initial_condition}) give
$\Psi_1(x,t)=\Psi_2(-x,t)$ for the numerical evolution.
We therefore focus on the $\Psi_1$ component below.

For repulsive interactions ($\sigma=-1$), the parameters
marked by the triangles in Fig.~\ref{fig:repulsive_MI} make $\omega$
and $2\omega$ unstable, while $3\omega$ is stable.
Thus, $n=2$, and we compare the simulation with an exact
$M=2$ multi-AB solution containing the modulation
wavenumbers $\{\omega,2\omega\}$.

{Figs.~\ref{fig:repulsive_multiAB}(a) and~\ref{fig:repulsive_multiAB}(c) show that the exact
solution reproduces the numerical two-breather localization
sequence and spatial modulation despite the different
initial profiles.}
Fig.~\ref{fig:repulsive_multiAB}(b) shows the accompanying redistribution
of power among successive sidebands.
Near the individual breather localization times,
Fig.~\ref{fig:repulsive_multiAB}(d) compares the numerical spectra with
the corresponding elementary-AB spectra.
Together, these comparisons support the description of
the selected two-breather event using the unstable
harmonics identified in the MI phase diagram.

{For attractive interactions ($\sigma=+1$), varying SOC
at fixed carrier wavenumber changes what harmonics
and AB branches are unstable.}
The two examples marked in Fig.~\ref{fig:attractive_MI} illustrate
multiple unstable branches at one harmonic and frequency
hopping across a stable gap, respectively.

For $k_L=0.6$ in Fig.~\ref{fig:attractive_MI}(b$_1$), the fundamental
$\omega$ supports two unstable branches with equal growth
rates and opposite modulation phase velocities.
The harmonics $2\omega$ and $3\omega$ each support one
unstable branch with zero modulation phase velocity.
Thus, $n=3$, while the selected exact solution has
$M=4$ constituents with modulation wavenumbers
$\{\omega,\omega,2\omega,3\omega\}$.
The repeated $\omega$ represents the two distinct
branches at the fundamental.

{Figure~\ref{fig:attractive_multiAB} shows that this exact four-AB solution
reproduces the breather localization sequence, spatial
modulation, and sideband power redistribution observed
in numerical simulations.}
Panel (d) compares the numerical spectra near the
constituent localization times with Fourier projections
of the complete multi-AB solution.

Increasing $k_L$ from $0.6$ to $1$ at fixed $k=0.2$
and $\omega=1$ moves $2\omega$ into the stability gap,
while $\omega$ and $3\omega$ remain unstable
[Fig.~\ref{fig:attractive_MI}(b$_2$)].
The fundamental supports two unstable branches and
$3\omega$ supports one.
Thus, $n=3$, and the selected exact solution has $M=3$
constituents with modulation wavenumbers
$\{\omega,\omega,3\omega\}$.
{The stable harmonic $2\omega$ is not an independent MI breather
channel, but the nonlinear mixing can still generate
spectral power at this harmonic.}

{Figure~\ref{fig:frequency_hopping} shows that the selected three-AB solution
reproduces the numerical frequency-hopping sequence,
spatial modulation, and spectral redistribution.}
Panel (d) compares the numerical spectra with Fourier
projections of the complete multi-AB solution.
Fourier amplitudes below $0.002$ are hidden only in
panel (b), so weak second-harmonic components may be
present even when they are not visible.

These zero-Raman examples exhibit FPUT-type recurrence
within the displayed time intervals.
They show how SOC changes the participating harmonics
and unstable branches, leading to different breather
sequences and spectral redistribution.

\subsection{Nonlinear Evolution with Finite Raman Coupling}

{We next examine how finite Raman coupling affects
unstable harmonics and their nonlinear spectral evolution,}
building on earlier studies of MI dynamics in SOC
BECs~\cite{Mithun2019}. We use direct simulations to connect
the harmonic instability map in Fig.~\ref{fig:finite_Raman_MI}
with the subsequent nonlinear dynamics.

{We apply the cosine modulation in
Eq.~(\ref{initial_condition}) to the in-phase background
in Eq.~(\ref{finite_Raman_background}), taking
$\Omega_R=0.26$, $k_L=1.2$, $k=0$,
$\sigma=N=\omega=1$, and $\epsilon=0.002$.}
These parameters correspond to the gain profile in
Fig.~\ref{fig:finite_Raman_MI}(b) and the associated
point in Fig.~\ref{fig:finite_Raman_MI}(c).

{The first three harmonics are linearly unstable,
while the fourth is stable.} In particular, $2\omega$
exhibits oscillatory MI, whereas it lies in the stable gap
in the zero-Raman frequency-hopping example of Fig.~\ref{fig:frequency_hopping}.
The two examples use different carrier wavenumbers and Raman couplings;
the zero-Raman reduction to $b=k+k_L$ does not apply
at finite Raman coupling.

\begin{figure}[th]
\centering
\includegraphics[width=\columnwidth]{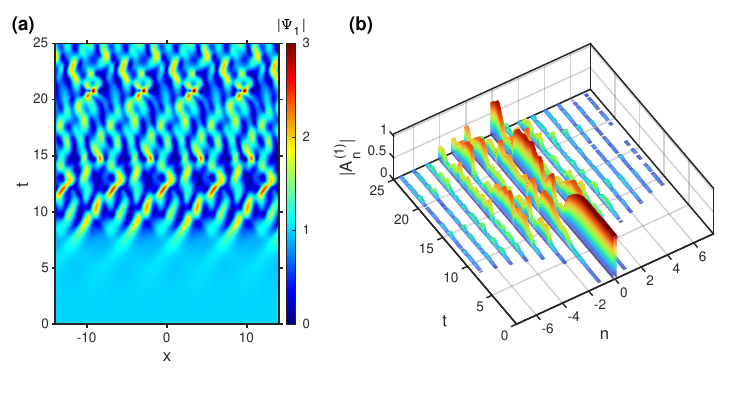}
\vspace{-2mm}
\caption{Finite-Raman nonlinear evolution for the attractive
interaction ($\protect\sigma=1$) at $\Omega_R=0.26$, $k=0$, $k_L=1.2$, and
$N=\protect\omega=1$, with $\epsilon=0.002$. (a) Direct numerical evolution
of $|\Psi_1|$ for $0\leq t\leq25$. (b) Corresponding discrete
spatial-Fourier amplitudes of $\Psi_1$. Amplitudes below
$2\times10^{-3}$ are suppressed only for visualization in panel (b).}
\label{fig:finite_Raman_dynamics}
\end{figure}

Fig.~\ref{fig:finite_Raman_dynamics} shows the corresponding
direct numerical evolution. {Within the displayed interval, the field develops
repeated localized events together with sustained occupation of several
Fourier sidebands.} {In particular, the second harmonic is now an independently
unstable linear channel, in contrast to its role as a harmonic lying in the stability gap in the zero-Raman
example displayed in Fig.~\ref{fig:frequency_hopping}.} The resulting evolution therefore illustrates the
nonlinear consequence of the Raman-induced rearrangement of the accessible MI
channels.

{The finite-Raman example shows that the Raman coupling can
rearrange the unstable harmonic channels and thereby change
the subsequent spectral evolution.} {In particular, a harmonic
that is stable in the zero-Raman frequency-hopping regime
becomes independently unstable, as a linear MI channel, at finite Raman
coupling.} {This comparison clarifies how the Raman coupling
modifies the relation between the linear MI spectrum and the
nonlinear sideband dynamics.}

{The finite-Raman system is not integrable, therefore we do not
relate the localized events, observed in Fig.~\ref{fig:finite_Raman_dynamics}, to an
exact multi-AB decomposition.} We therefore turn to the
integrable zero-Raman limit, where the exact AB spectrum
allows the gain, carrier depletion, sideband asymmetry, and
higher-order spectral redistribution to be analyzed
quantitatively.

\section{{EXACT AB SPECTRA AND BRANCH-RESOLVED SPECTRAL REDISTRIBUTION}}

\subsection{{The exact Fourier Spectrum and Spectral Characteristics}}

We now derive the exact Fourier spectrum of an elementary AB
in the integrable zero-Raman limit. {The zero-Raman model can be
mapped to the Manakov system by the gauge transformation, but the
transformation back to the original SOC-BEC fields restores
SOC-dependent carrier phases and effective wave numbers.} {The
Manakov spectrum therefore does not directly give the complete
laboratory-frame spectrum of the SOC-BEC system.}

{The Fourier spectrum of the
Manakov AB was produced in Ref.~\cite{Chen2021}, but the positive-order coefficients were expressed
through order-dependent derivatives.} {This form does not provide
a uniform closed expression that directly exposes the decay of
arbitrary higher-order sidebands.} {Here we overcome this
limitation by using an exterior-pole representation for the
positive-order coefficients, together with the interior-pole
representation for the negative-order ones.} {The
resulting spectrum is explicit for every sideband order and
retains the SOC dependence after transforming back to the
original fields.}

For the counterpropagating zero-Raman background,
$b_1=-b_2=b=k+k_L$. {Thus, at a fixed carrier wavenumber $k$, the
SOC strength changes the spectrum relative to the carrier through the
effective background wavenumber $b$. At fixed $b$, SOC changes
the carrier phases but does not change the envelope spectrum itself.}

Writing the elementary AB as
\begin{equation}
\begin{aligned}\label{new_form_AB_solution}
\Psi_j(x,t)=\Psi_{j0}(x,t)\mathcal{R}_j(x,t),\qquad j=1,2,
\end{aligned}
\end{equation}
the normalized complex Fourier coefficients are
\begin{equation}
\begin{aligned}\label{Fourier_spectrum}
A_n^{(j)}(t)=\frac{1}{L}\int_0^L
\mathcal{R}_j(x,t)e^{-in\omega x}\,dx .
\end{aligned}
\end{equation}
Here $n$ denotes the sideband order. The residue calculation
based on the exterior- and interior-pole representations gives
the exact coefficients
\begin{equation}
A_n^{(j)}(t)=-2i\tau_I\mathcal{L}_1^{(j)}(t)
\mathbb{Z}_1(t)^{|n|},\qquad n\leq-1,
\label{negative_sidebands}
\end{equation}
\begin{equation}
A_n^{(j)}(t)=-2i\tau_I\mathcal{L}_2^{(j)}(t)
\mathbb{Z}_2(t)^{-n},\qquad n\geq1,
\label{positive_sidebands}
\end{equation}
and
\begin{equation}
A_0^{(j)}(t)=1-2i\tau_I\left[\mathcal{L}_1^{(j)}(t)
+\frac{\mathcal{P}_3^{(j)}(t)}{\mathcal{Q}_2^*(t)}\right].
\label{pump_mode}
\end{equation}
The definitions of $\mathbb{Z}_{1,2}$ and
$\mathcal{L}_{1,2}^{(j)}$ are given in Appendix~\ref{app:exact_spectrum}.
{The exterior-pole form of the positive-order coefficients removes
the order-dependent derivatives used in Ref.~\cite{Chen2021}.}
{These formulas provide the exact complex coefficient of every
member of the infinite sideband sequence.} The corresponding
normalized power spectrum is
{\begin{equation}
\begin{aligned}
|A_{\pm n}^{(j)}(t)|^2
&=\frac{S_j(t)}{2}\left[1\pm a_j(t)\right]
\left[1-q(t)\right]q(t)^{n-1},\qquad n\geq1 .
\end{aligned}
\label{geometric_power_spectrum}
\end{equation}}

The three quantities in Eq.~(\ref{geometric_power_spectrum}) are
defined by
{\begin{equation}
\begin{aligned}
S_j(t)&\equiv\sum_{n\neq0}|A_n^{(j)}(t)|^2
=1-|A_0^{(j)}(t)|^2,\\
a_j(t)&\equiv
\frac{\displaystyle\sum_{n=1}^{\infty}
\left[|A_n^{(j)}(t)|^2-|A_{-n}^{(j)}(t)|^2\right]}
{S_j(t)},\\
q(t)&\equiv|\mathbb{Z}_1(t)|^2=|\mathbb{Z}_2(t)|^{-2}.
\end{aligned}
\label{spectral_quantities}
\end{equation}}
{Here $S_j$ is the total power transferred from the carrier,
$a_j$ measures the imbalance between the two spectral
directions, and $q$ gives the power ratio between successive
sideband orders.} {Equation~(\ref{geometric_power_spectrum})
therefore provides an exact description of the carrier depletion,
directional sideband imbalance, and the geometric decay of all
higher-order sidebands.}

\begin{figure*}[!t]
\centering
\includegraphics[width=\textwidth]{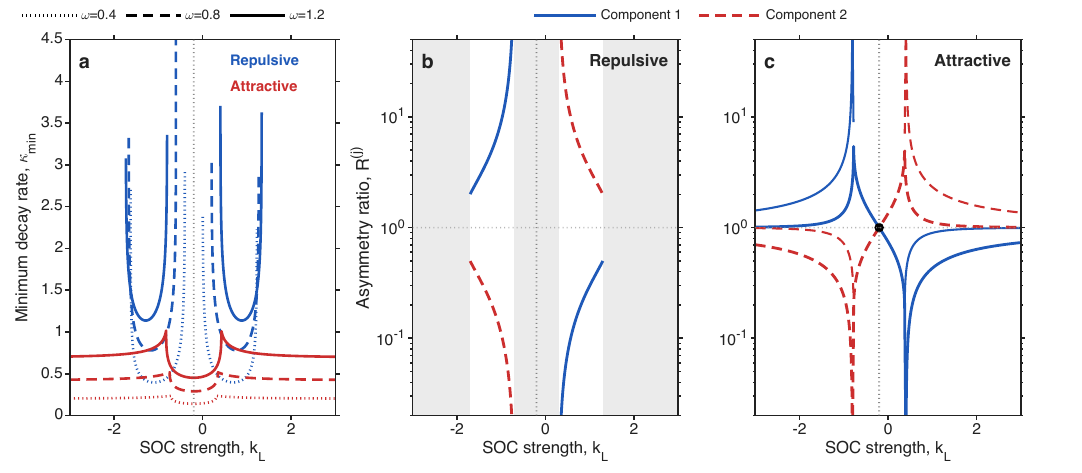}\vspace{-2mm}
\caption{{Analytical spectral characteristics versus the SOC parameter $k_L$ for $N=1$
and $k=0.2$. (a) Minimum decay rate $\protect\kappa_{\min}$ for
$\protect\omega=0.4,0.8,1.2$. (b,c) Center-time asymmetry ratio $R^{(j)}(0)$ at
$\protect\omega=1$ for the repulsive and attractive interactions. Blue and
red denote the two interaction signs in (a) and components 1 and 2 in (b,c),
with line styles distinguishing the remaining cases. Gray regions in (b)
mark fundamental-stability gaps; the black point in (c) marks
$b=k+k_L=0$ and $R^{(j)}=1$. Only the fastest-growing regular branch or
branches are shown.}}
\label{fig:spectral_characteristics}
\end{figure*}

For Fig.~\ref{fig:spectral_characteristics}, we also use
{\begin{equation}
\begin{aligned}
\kappa(t)&=-\frac{1}{2}\ln q(t),\qquad
R^{(j)}(t)=\sqrt{\frac{1+a_j(t)}{1-a_j(t)}} .
\end{aligned}
\label{spectral_characteristic_relations}
\end{equation}}
At the AB center, $q$ is maximal and the spectrum is broadest,
so $\kappa_{\min}=\kappa(0)$. The quantities
$\kappa_{\min}$ and $R^{(j)}(0)$ characterize the minimum
sideband decay and the center-time directional imbalance,
respectively.

\subsection{{Gain--Depletion Ordering in Coexisting AB Branches}}

The MI gain describes the initial growth of a weak modulation,
whereas the carrier depletion measures the nonlinear power
transfer during a complete AB event. {Comparing these two
quantities tests whether the linear MI gain can predict the
strength and spectral distribution of the subsequent nonlinear
evolution.}

For attractive interactions ($\sigma=+1$), two regular
elementary AB branches coexist in the following open domain
{\begin{equation}
0<\omega^2<2N,\qquad
\frac{|\omega|}{2}<|b|<
\frac{N}{\sqrt{4N-\omega^2}} .
\label{branch_coexistence}
\end{equation}}
{We compare the two branches at the same background parameters
$N$ and $b$ and the same modulation wavenumber $\omega>0$.
The event centers are aligned at $t=0$.} We denote the branches
by $H$ and $L$ according to their MI gains,
\begin{equation*}
G_\alpha=|\omega\operatorname{Im}\tau_\alpha|,
\qquad \alpha=H,L .
\end{equation*}
Their exact spectra satisfy
{\begin{equation}
\begin{aligned}
&G_H>G_L>0,\qquad
S_{j,H}(0)<S_{j,L}(0),\\
&q_H(0)>q_L(0),\qquad
|a_{j,H}(0)|<|a_{j,L}(0)|,\quad j=1,2 .
\end{aligned}
\label{gain_transfer_ordering}
\end{equation}}

{For each branch, the carrier depletion is maximal at the event
center, hence $S_{j,\alpha}(0)$ is the maximum depletion during
the complete AB event.} {The higher-gain branch therefore produces a smaller maximum carrier
depletion, although its high-order sidebands decay more slowly
and are more evenly distributed between the two spectral
directions.} {Thus, the linear MI gain does not determine
the maximum carrier depletion and the total transferred sideband
power.} This result shows why the nonlinear spectral properties
of coexisting AB branches must be analyzed beyond the linear
gain spectrum.

\begin{figure}[!t]
\centering
\includegraphics[width=\columnwidth]
{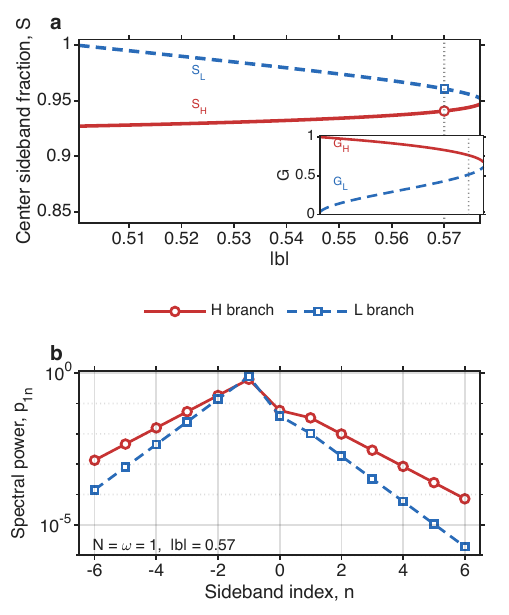}\vspace{-2mm}
\caption{Gain and carrier depletion for the two coexisting
AB branches with attractive interactions and $N=\omega=1$.
(a) Maximum carrier depletions $S_{j,H/L}(0)$ versus $|b|$;
the inset shows the corresponding MI gains $G_{H/L}$.
(b) Component-1 power spectra
$p_{1n}=|A_n^{(1)}(0)|^2$ at the event center for $b=0.57$.}
\label{fig:gain_depletion}
\end{figure}

Figure~\ref{fig:gain_depletion}(a) illustrates the opposite ordering of the
MI gain and maximum carrier depletion throughout the
coexistence domain: the higher-gain branch transfers less
power from the carrier. At $|b|=N/\sqrt{4N-\omega^2}$,
the two AB branches coalesce, explaining the convergence
of the depletion curves and the gain curves in
Fig.~\ref{fig:gain_depletion}(a). Fig.~\ref{fig:gain_depletion}(b) shows that its
high-order sidebands nevertheless decay more slowly and
have weaker directional imbalance. {A broader sideband
distribution therefore does not imply greater total
carrier depletion.}

\subsection{{High-Order Sideband Decay and Directional Asymmetry}}

{The exact Fourier coefficients separate two distinct
characteristics of an elementary-AB spectrum: the decay of power with sideband
order and the imbalance between positive- and negative-order sidebands. The
former is governed by the pole modulus, or equivalently by $q$ and $\kappa$,
whereas the latter is determined by the component- and branch-dependent
residue amplitudes.}

\subsubsection{{Geometric Sideband Decay}}

{The sideband-order dependence of the exact coefficients is
geometric. For a regular elementary AB, the negative- and positive-order
sidebands contain the factors $\mathbb{Z}_{1}^{|n|}$ and
$\mathbb{Z}_{2}^{-n}$, respectively. Their residue prefactors may differ, but
$|\mathbb{Z}_{1}|<1<|\mathbb{Z}_{2}|$ and
$|\mathbb{Z}_{1}\mathbb{Z}_{2}|=1$ imply the same exponential decay rate for
both sets of sidebands:}
{{\begin{equation}
\begin{aligned}
|A_{\pm n}^{(j)}(t)|&\propto e^{-\kappa(t)n},\qquad n\geq1,\\
\kappa(t)&=-\ln|\mathbb{Z}_{1}(t)|
=\ln|\mathbb{Z}_{2}(t)|>0 .
\end{aligned}
\end{equation}}%
}
{Smaller $\kappa$ therefore corresponds to a broader
distribution over high-order sidebands. At the AB center, $\Theta=0$, the
spectrum is broadest and $\kappa$ reaches its minimum value
$\kappa_{\min}$. The pole modulus then gives}
{\begin{equation}
|\tau_I|=\frac{|\omega|}{2\sinh(\kappa_{\min})}.
\label{eq_binding}
\end{equation}%
}
{At fixed nonzero $\omega$, Eq.~(\ref{eq_binding}) shows that
a larger MI gain corresponds to a smaller $\kappa_{\min}$ and hence to a
broader conditional high-order sideband distribution. {As emphasized in
Sec.~IV\,B, this relation defines how the transferred power is distributed
over sideband order and does not determine the total carrier depletion
$S_j$.}}

{Fig.~\ref{fig:spectral_characteristics}(a) shows $\kappa_{\min}$ as $k_L$ is
varied at fixed $k$. In the zero-Raman problem this is equivalently a scan of
the effective background wavenumber $b=k+k_L$, which explains the symmetry
about $k_L=-k$.}
Varying the SOC parameter $k_L$ at fixed carrier wavenumber
$k$ therefore moves the prepared state through different
admissible branches and changes the corresponding spectrum relative to the carrier.
{For the repulsive interaction, approaching an MI boundary gives
$G\to0$ and $\kappa_{\min}\to\infty$. For the attractive interaction, the
cusplike points $k_L^c=-k\pm N/\sqrt{4N-\omega^2}$ occur at $\Delta=0$,
where the relevant branch roots meet and the regular branch represented in
Fig.~\ref{fig:spectral_characteristics}(a) changes. The comparison with Fig.~\ref{fig:gain_depletion} further shows
why the high-order tail width must be distinguished from the total carrier
depletion.}

\subsubsection{{Directional Asymmetry and Branch Symmetry}}

{Although positive- and negative-order sidebands have the same
geometric decay rate, their amplitudes need not be equal. For a regular
elementary AB, their amplitude ratio is independent of the sideband order
$n$:}
{\begin{equation}
R^{(j)}(t)=\left|\frac{A_n^{(j)}(t)}{A_{-n}^{(j)}(t)}\right|
=\left|\frac{\mathcal{L}_{2}^{(j)}(t)}
{\mathcal{L}_{1}^{(j)}(t)}\right|,\qquad n\geq1 .
\end{equation}%
}
The directional asymmetry is determined by the branch
eigenvalue and the effective background wavenumber through the residue
amplitudes. For the counterpropagating zero-Raman background,
$b_j=\pm(k+k_L)$ and $b_1=-b_2$, so the corresponding symmetry-related
branches satisfy
$R^{(1)}(k_L)=R^{(2)}(-k_L-2k)$.

In the $\Delta<0$ region, the two nondegenerate branches have
equal MI gains and satisfy
$\tau_b=-\omega-\tau_a^*$. Component exchange maps branch $a$
to branch $b$ and vice versa, which gives
\begin{equation*}
R_a^{(1)}R_b^{(2)}=1,
\qquad
R_b^{(1)}R_a^{(2)}=1 .
\end{equation*}
For a stationary-modulation branch that is invariant under this symmetry,
the relation reduces to $R^{(1)}R^{(2)}=1$. At $b=0$, i.e.,
$k_L=-k$, the two components are symmetry-related and the
elementary-AB spectrum is symmetric, with $R^{(j)}=1$.

\begin{figure}[!t]
\centering
\includegraphics[width=\columnwidth]
{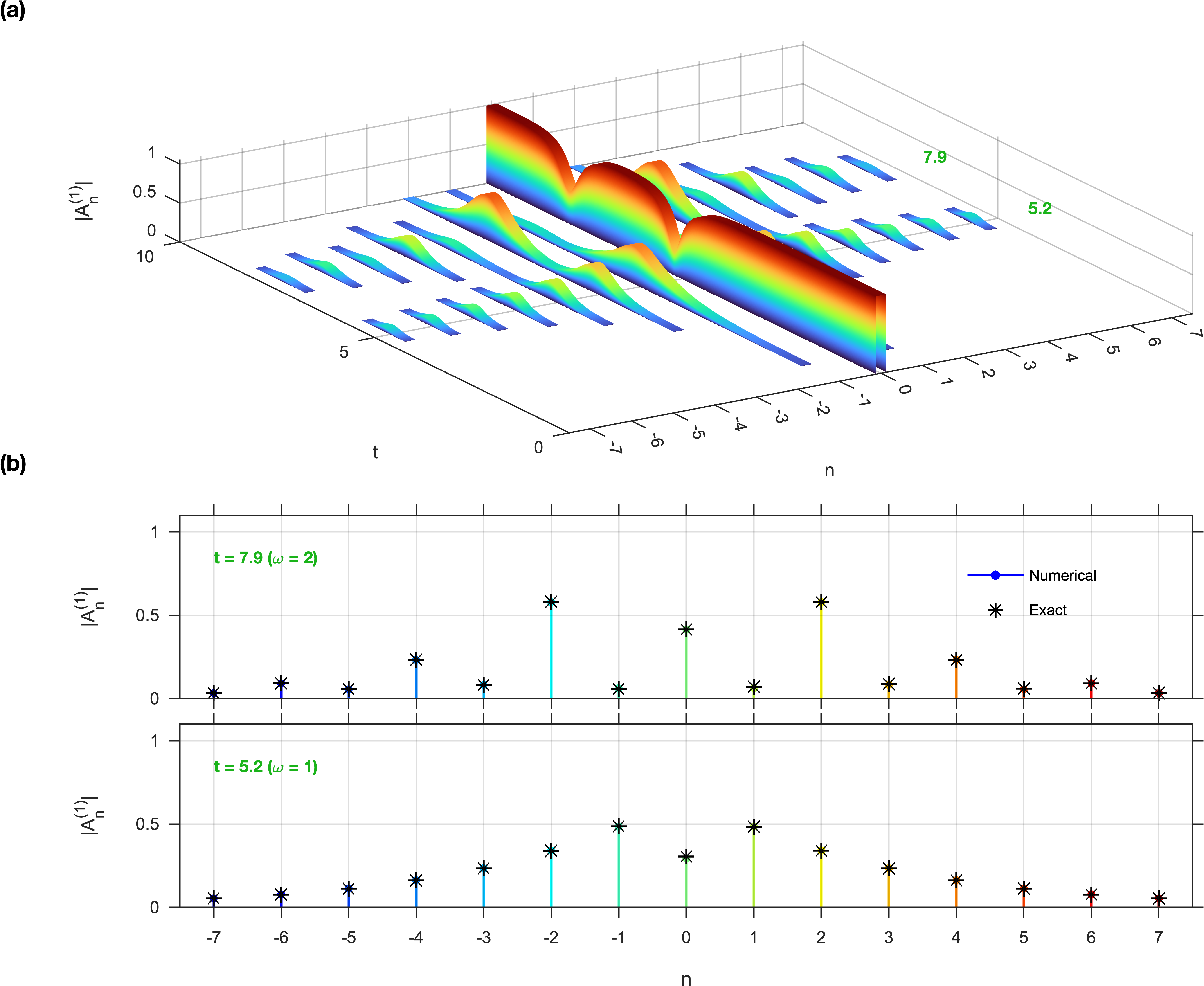}\vspace{-2mm}
\caption{{The symmetric breather spectrum for the attractive
interaction at $b=k+k_L=0$, with $k=-0.2$, $k_L=0.2$, and
$\omega=1$. The positive- and negative-order sidebands have
equal powers, showing that the SOC-induced directional
asymmetry vanishes when the effective shifted wavenumber is zero.}}
\label{fig:symmetric_breather}
\end{figure}

{The symmetric spectrum is illustrated in Fig.~\ref{fig:symmetric_breather}. For
$k=-0.2$, $k_L=0.2$, and $\omega=1$, the positive- and
negative-order sidebands have equal powers, consistent with
$R^{(j)}=1$ for both components.}

This example directly confirms the symmetry relation obtained
above: once $b$ vanishes, the two spectral directions carry equal
power in each component.

The exact coefficients also identify two special center-time
configurations. {Complete depletion of component $j$ occurs when
$S_j(0)=1$, equivalently $A_0^{(j)}(0)=0$. A one-sided spectrum
occurs when $a_j(0)=\pm1$, in which case all sidebands vanish in one
spectral direction, and $R^{(j)}(0)$ tends to infinity or
zero.} These configurations are special consequences of the
general exact spectrum and must still satisfy the characteristic
equation and the regularity conditions of the elementary AB.

{Figs.~\ref{fig:spectral_characteristics}(b) and~\ref{fig:spectral_characteristics}(c) summarize these
center-time asymmetry relations. {In the gray intervals of Fig.~\ref{fig:spectral_characteristics}(b),
no finite-gain regular AB exists for the plotted branch, so the corresponding
AB asymmetry ratio is undefined.}}
{This does not imply that the same Fourier harmonic must vanish in
the case of the general nonlinear evolution, because nonlinear mixing may still populate
that harmonic.}
Fig.~\ref{fig:spectral_characteristics}(c) likewise should not be interpreted as
following one AB branch continuously over the entire $k_L$ range: the outer
$\Delta<0$ regions contain the two equal-gain nondegenerate drifting branches,
whereas the central $\Delta>0$ region displays the fastest-growing regular
stationary-modulation branch.

\section{Conclusions}

We investigated higher-order MI and nonlinear spectral redistribution
in a spin-orbit-coupled binary BEC with equal interactions. {For finite
Raman coupling, the system is nonintegrable, and the MI spectrum
contains distinct nonoscillatory and oscillatory instability regimes.}
Raman coupling changes the connectivity of the MI bands and the
accessible harmonic channels. Direct simulations confirm that this
rearrangement modifies the subsequent nonlinear sideband dynamics.

In the zero-Raman limit, the equations are gauge-related to the
Manakov system, while the inverse transformation restores
SOC-dependent phases and effective wave numbers. {At fixed carrier
wave number $k$, varying the SOC strength changes
$b=k+k_L$ and shifts successive harmonics across MI bands and
stability gaps.} This provides control over higher-order MI, AB
branches, and frequency-hopping sequences. Exact multi-AB solutions
and numerical simulations further demonstrate the resulting coherent
evolution and FPUT-type recurrence.

{Starting from the known Manakov AB spectrum, we derived exact
algebraic coefficients for the carrier and arbitrary-order sidebands
without order-dependent differentiation.} {The resulting geometric
power spectrum covers the complete infinite sideband sequence and
provides a unified description of carrier depletion, directional
imbalance, and spectral decay.} {We have also established the exact relations
between MI gain and spectral decay, showing that the higher-gain branch can undergo weaker maximum carrier depletion,
while its high-order sidebands decay more slowly and show weaker
directional imbalance.}

The effective background wavenumber $b$ determines the spectrum relative
to the carrier, while $k$ and $k_L$ provide distinct laboratory controls.
{In particular, $k_L=-k$ can produce a symmetric spectrum with
$b=0$ even when both parameters are nonzero.} The multi-AB results
describe coherent states generated by specified periodic
perturbations; broadband-noise-driven dynamics may involve additional
nonlinear processes.

\begin{acknowledgments}
This work was supported by the National Natural Science Foundation of China
(Grant Nos. 12505006, 12675003, 11975172 and 12381240286).
\end{acknowledgments}

\appendix

\section{{Exact Multi-AB Solutions in the Integrable SOC-BEC Limit}}

In the integrable limit considered in this work, namely $V(x)=0$ and
{$\Omega_R=0$}, the SOC-BEC model based on Eq.~(\ref{model_equation}) admits
exact multi-AB solutions. {Here $M$ denotes the number of breather constituents,
rather than the highest unstable harmonic index $n$ defined in Sec.~II\,D.
Each constituent $s=1,\ldots,M$ is specified by a modulation wavenumber
$\omega_s$ and an admissible nonreal eigenvalue $\tau_s$ satisfying
Eq.~(\ref{characteristic_equation}) at $\omega=\omega_s$, together with the
regularity conditions stated after Eq.~(\ref{tau}). We define
$\widetilde{\tau}_s\equiv\tau_s+\omega_s$ and use $x_{s0}$ and $t_{s0}$
for its spatial and temporal translations. The component background
wavenumbers are $b_1=-b_2=b=k+k_L$. The construction includes the elementary
case $M=1$.}

{For completeness, the auxiliary parameters entering the elementary AB in
Eq.~(\ref{breather}) are given here. With
$\widetilde{\tau}\equiv\tau+\omega$,
\begin{equation}
\begin{aligned}\label{parameters2}
&\vartheta_j=\frac{\vartheta_{1j}-\vartheta_{2j}}{2},\qquad
\upsilon_j=\log\frac{|\tau+b_j|}{|\widetilde{\tau}+b_j|},\\
&\varpi=\left|\frac{2\tau_I}{2\tau_I-i\omega}\right|,\qquad
\theta_{1j}=\frac{\vartheta_{1j}+\vartheta_{2j}}{2},\\
&\theta_{2j}=\arg\frac{\tau^*+b_j}{\widetilde{\tau}+b_j},\\
&\vartheta_{1j}=\arg\frac{\tau^*+b_j}{\tau+b_j},\qquad
\vartheta_{2j}=\arg\frac{\widetilde{\tau}^*+b_j}{\widetilde{\tau}+b_j}.
\end{aligned}
\end{equation}}

For an $M$-breather solution, we write
\begin{equation}
\begin{aligned}\label{multi_AB}
\Psi_j[M]=\Psi_{j0}\frac{\det(\boldsymbol{M}^{(j)})}{\det(\boldsymbol{M})},%
\qquad j=1,2, \end{aligned}
\end{equation}%
where $\Psi _{j0}$ is the plane-wave background, see
Eq.~(\ref{plane_wave_solution}). The $M\times M$ matrices
$\boldsymbol{M}$ and $\boldsymbol{M}^{(j)}$ are defined as
\begin{equation}
\begin{aligned}\label{matrix_M}
&{\boldsymbol{M}=\left(m_{r,s}\right)_{1\leq r,s\leq M}},\\
&{\boldsymbol{M}^{(j)}=\left(m_{r,s}^{(j)}\right)_{1\leq r,s\leq M}},
\end{aligned}
\end{equation}%
with matrix elements
\begin{equation}
\begin{aligned}\label{matrix_elements}
&{m_{r,s}=\boldsymbol{u}(s)\mathbb{M}(r,s)\boldsymbol{v}(r)},\\
&{m_{r,s}^{(j)}=\boldsymbol{u}(s)\widetilde{\mathbb{M}}_j(r,s)
\boldsymbol{v}(r)}.
\end{aligned}
\end{equation}%
{The vectors $\boldsymbol{u}(s)$ and $\boldsymbol{v}(r)$, together with the
matrices $\mathbb{M}(r,s)$ and $\widetilde{\mathbb{M}}_{j}(r,s)$, are given by}
\begin{equation}
\begin{aligned}\label{vectors_matrices}
&{\boldsymbol{u}(s)=(\varphi_s^*,\widetilde{\varphi}_s^*),\qquad
\boldsymbol{v}(r)=(\varphi_r,\widetilde{\varphi}_r)^T},\\
&{\mathbb{M}(r,s)=
\begin{pmatrix}
\frac{1}{\tau_s^*-\tau_r} &
\frac{1}{\tau_s^*-\widetilde{\tau}_r}\\
\frac{1}{\widetilde{\tau}_s^*-\tau_r} &
\frac{1}{\widetilde{\tau}_s^*-\widetilde{\tau}_r}
\end{pmatrix}},\\
&{\widetilde{\mathbb{M}}_j(r,s)
=\boldsymbol{D}_{j,s}^{(+)}\,\mathbb{M}(r,s)\,\boldsymbol{D}_{j,r}^{(-)}},\\
&{\boldsymbol{D}_{j,s}^{(+)}
=\mathrm{diag}(\tau_s^*+b_j,\widetilde{\tau}_s^*+b_j)},\\
&{\boldsymbol{D}_{j,r}^{(-)}
=\mathrm{diag}\left(\frac{1}{\tau_r+b_j},
\frac{1}{\widetilde{\tau}_r+b_j}\right)}.
\end{aligned}
\end{equation}%
{The associated eigenfunctions are}
\begin{equation}
\begin{aligned}\label{eigenfunctions}
&{\varphi_s(x,t)=\exp\left\{i\tau_s\left[(x-x_{s0})+\frac{1}{2}
\tau_s(t-t_{s0})\right]\right\}},\\
&{\widetilde{\varphi}_s(x,t)=\exp\left\{i\widetilde{\tau}_s
\left[(x-x_{s0})+\frac{1}{2}\widetilde{\tau}_s(t-t_{s0})\right]\right\}}.
\end{aligned}
\end{equation}

\vspace{12mm}

\section{{Derivation of the Exact Fourier Spectrum}}
\label{app:exact_spectrum}

The periodic modulation in Eq.~(\ref{new_form_AB_solution}) has the
rational form
\begin{equation}
\begin{aligned}
\mathcal{R}_j
&=1-2i\tau_I
\frac{\mathcal{P}_1^{(j)}
+\mathcal{P}_2^{(j)}e^{i\omega x}
+\mathcal{P}_3^{(j)}e^{-i\omega x}}
{\mathcal{Q}_1
+\mathcal{Q}_2e^{i\omega x}
+\mathcal{Q}_2^*e^{-i\omega x}} .
\end{aligned}
\label{modulation_factor}
\end{equation}
The time-dependent coefficients are
\begin{equation}
\begin{aligned}
\mathcal{P}_1^{(j)}
&=\frac{e^\Theta}{\tau+b_j}
+\frac{e^{-\Theta}}{\tau+b_j+\omega},\\
\mathcal{P}_2^{(j)}
&=\frac{e^{i\Phi}}{\tau+b_j+\omega},
\qquad
\mathcal{P}_3^{(j)}
=\frac{e^{-i\Phi}}{\tau+b_j},\\
\mathcal{Q}_1&=2\cosh\Theta,\qquad
\mathcal{Q}_2
=\frac{2\tau_I}{2\tau_I-i\omega}e^{i\Phi},
\end{aligned}
\label{t_dependent_coefficients}
\end{equation}
where $\Theta=\omega\tau_I t$,
$\Phi=\omega(\tau_R+\omega/2)t$, and
$b_1=-b_2=b=k+k_L$.

Setting $\mathbb{Z}=e^{i\omega x}$ transforms the Fourier integral
into a contour integral around the unit circle. The denominator
polynomial is
$\mathcal{Q}_2\mathbb{Z}^2+
\mathcal{Q}_1\mathbb{Z}+\mathcal{Q}_2^*$, with roots
\begin{equation}
\mathbb{Z}_{1,2}
=\frac{-\mathcal{Q}_1\pm\mathcal{D}}{2\mathcal{Q}_2},
\qquad
\mathcal{D}=2\sqrt{\cosh^2\Theta-\varpi^2},
\label{denominator_roots2}
\end{equation}
where
$\varpi=2|\tau_I|/\sqrt{4\tau_I^2+\omega^2}<1$.
For a regular AB,
$|\mathbb{Z}_1|<1<|\mathbb{Z}_2|$ and
$|\mathbb{Z}_1\mathbb{Z}_2|=1$.

Define
\begin{equation}
\mathcal{L}_{1,2}^{(j)}
=\frac{
\mathcal{P}_1^{(j)}
+\mathcal{P}_2^{(j)}\mathbb{Z}_{1,2}
+\mathcal{P}_3^{(j)}\mathbb{Z}_{1,2}^{-1}}
{\mathcal{D}} .
\label{L_coeficients2}
\end{equation}
For $n\leq-1$, the origin is regular and the interior pole
$\mathbb{Z}_1$ gives Eq.~(\ref{negative_sidebands}).
{For $n\geq1$, an interior-contour calculation contains an
order-dependent pole at the origin. Instead, evaluating the contour from
the exterior pole $\mathbb{Z}_2$ leads to
Eq.~(\ref{positive_sidebands}) directly, without derivatives
whose order grows with $n$.} For $n=0$, the origin and the
interior pole give Eq.~(\ref{pump_mode}).

The equivalent carrier expression
\begin{equation*}
A_0^{(j)}
=1-2i\tau_I
\left[
\mathcal{L}_2^{(j)}
+\frac{\mathcal{P}_2^{(j)}}{\mathcal{Q}_2}
\right]
\end{equation*}
provides an algebraic consistency check.

Component-norm conservation and the unit-modulus temporal far
field imply
\begin{equation*}
\frac{1}{L}\int_0^L|\mathcal{R}_j(x,t)|^2\,dx=1.
\end{equation*}
Parseval's identity therefore gives
$\sum_{n=-\infty}^{\infty}|A_n^{(j)}|^2=1$, which yields
$S_j=1-|A_0^{(j)}|^2$. {Equations
(\ref{negative_sidebands}) and (\ref{positive_sidebands})
then give the geometric power spectrum in
Eq.~(\ref{geometric_power_spectrum}).}

\section{Proof of the Gain--Depletion Ordering}
\label{app:gain_depletion}

For the two branches in Eq.~(\ref{branch_coexistence}), define
\begin{equation*}
r_\alpha=
\sqrt{\frac{\omega^2}{4}
+\left[\operatorname{Im}\tau_\alpha\right]^2},
\qquad \alpha=H,L .
\end{equation*}
The coexistence conditions imply
$r_H>r_L>|b|>|\omega|/2$. At the event center, the exact
coefficients give
\begin{equation*}
q_\alpha(0)=
\frac{r_\alpha-|\omega|/2}{r_\alpha+|\omega|/2},
\qquad
|a_{j,\alpha}(0)|=
\frac{2|b|r_\alpha}{b^2+r_\alpha^2}.
\end{equation*}
The first expression increases with $r_\alpha$, whereas the
second decreases for $r_\alpha>|b|$. Together with
$\operatorname{Im}\tau_H>\operatorname{Im}\tau_L$, these
relations establish the gain, decay-factor, and imbalance
orderings in Eq.~(\ref{gain_transfer_ordering}).

For either branch, differentiation of the exact carrier
coefficient in Eq.~(\ref{pump_mode}) shows that
$S_{j,\alpha}(t)$ decreases with $|t|$. Thus,
$S_{j,\alpha}(0)$ is the maximum depletion during the
complete AB event. Direct substitution gives
\begin{equation}
\begin{aligned}
S_{j,L}(0)-S_{j,H}(0)
&=\frac{|\omega|(r_H-r_L)(N-\omega^2/2)^2}
{N[b^2+N-\omega^2/2+r_Hr_L]}\\
&\quad\times
\frac{1}{
r_Hr_L-b^2+\omega^2/2+
(|\omega|/2)(r_H+r_L)}>0,
\qquad j=1,2 .
\end{aligned}
\label{depletion_difference}
\end{equation}
{Every factor is positive in the open coexistence domain, proving
the opposite ordering of the MI gain and maximum carrier depletion.}
At the lower boundary $|b|=|\omega|/2$, the lower-gain branch
becomes singular and leaves the regular finite-gain AB family.
At the upper boundary
$|b|=N/\sqrt{4N-\omega^2}$, the two branches coalesce as
$\Delta\to0$, and their differences vanish.


\begin{thebibliography}{99}
\bibitem{Zakharov2009} V. E. Zakharov and L. A. Ostrovsky, Modulation
instability: The beginning, Physica D 238, 540 (2009).

\bibitem{Segur2007} H. Segur and D. M. Henderson, The modulation instability
revisited, Eur. Phys. J. Spec. Top. 147, 25 (2007).

\bibitem{Everitt2017} P. J. Everitt, M. A. Sooriyabandara, M. Guasoni, P. B.
Wigley, C. H. Wei, G. D. McDonald, K. S. Hardman, P. Manju, J. D. Close, C.
C. N. Kuhn, S. S. Szigeti, Y. S. Kivshar, and N. P. Robins, Observation of a
modulational instability in Bose-Einstein condensates, Phys. Rev. A 96,
041601 (2017).

\bibitem{Otajonov2022} S. R. Otajonov, E. N. Tsoy, and F. K. Abdullaev,
Modulational instability and quantum droplets in a two-dimensional
Bose-Einstein condensate, Phys. Rev. A 106, 033309 (2022).

\bibitem{Kraych2019} A. E. Kraych, P. Suret, G. El, and S. Randoux,
Nonlinear Evolution of the Locally Induced Modulational Instability in Fiber
Optics, Phys. Rev. Lett. 122, 054101 (2019).

\bibitem{Armaroli2023} A. Armaroli, G. Dujardin, A. Kudlinski, A. Mussot, S.
D. Bievre, and M. Conforti, Modulational instability in randomly
dispersion-managed fiber links, Phys. Rev. A 108, 023510 (2023).

\bibitem{Tabi2022} C. B. Tabi, H. Tagwo, C. G. Latchio Tiofack, and T. C.
Kofan\'{e}, Pure quartic modulational instability in weakly nonlocal
birefringent fibers, Opt. Lett. 47, 5557 (2022).

\bibitem{Kharif2010} C. Kharif, R. A. Kraenkel, M. A. Manna, and R. Thomas,
The modulational instability in deep water under the action of wind and
dissipation, Journal of Fluid Mechanics 664, 138 (2010).

\bibitem{Sun2025} W.-R. Sun and B. A. Malomed, Subharmonic modulational
instabilities, Physics Reports 1139, 1 (2025).

\bibitem{Perego2022} A. M. Perego, F. Bessin, and A. Mussot, Complexity of
modulation instability, Phys. Rev. Research 4, L022057 (2022).

\bibitem{Vanderhaegen2021} G. Vanderhaegen, C. Naveau, P. Szriftgiser, A.
Kudlinski, M. Conforti, A. Mussot, M. Onorato, S. Trillo, A. Chabchoub, and
N. Akhmediev, ``Extraordinary'' modulation instability in
optics and hydrodynamics, Proc. Natl. Acad. Sci. U.S.A. 118, e2019348118
(2021).

\bibitem{Ndebele2021} K. K. Ndebele, C. B. Tabi, C. G. L. Tiofack, and T. C.
Kofan\'{e}, Higher-order dispersion and nonlinear effects of optical fibers
under septic self-steepening and self-frequency shift, Phys. Rev. E 104,
044208 (2021).

\bibitem{Choudhuri2012} A. Choudhuri and K. Porsezian, Impact of dispersion
and non-Kerr nonlinearity on the modulational instability of the
higher-order nonlinear Schr\"odinger equation, Phys. Rev. A 85, 033820
(2012).

\bibitem{Gao2020} P. Gao, C. Liu, L.-C. Zhao, Z.-Y. Yang, and W.-L. Yang,
Modified linear stability analysis for quantitative dynamics of a perturbed
plane wave, Phys. Rev. E 102, 022207 (2020).

\bibitem{Kimmoun2016} O. Kimmoun, H. C. Hsu, H. Branger, M. S. Li, Y. Y.
Chen, C. Kharif, M. Onorato, E. J. R. Kelleher, B. Kibler, N. Akhmediev, and
A. Chabchoub, Modulation instability and phase-shifted Fermi-Pasta-Ulam
recurrence, Sci. Rep. 6, 28516 (2016).

\bibitem{Chin2015} S. A. Chin, O. A. Ashour, and M. R. Beli\'{c}, Anatomy of
the Akhmediev breather: Cascading instability, first formation time, and
Fermi-Pasta-Ulam recurrence, Phys. Rev. E 92, 063202 (2015).

\bibitem{Chen2021} S.-C. Chen, C. Liu, X. Yao, L.-C. Zhao, and N. Akhmediev,
Extreme spectral asymmetry of Akhmediev breathers and Fermi-Pasta-Ulam
recurrence in a Manakov system, Phys. Rev. E 104, 024215 (2021).

\bibitem{Pierangeli2018} D. Pierangeli, M. Flammini, L. Zhang, G. Marcucci,
A. J. Agranat, P. G. Grinevich, P. M. Santini, C. Conti, and E. DelRe,
Observation of Fermi-Pasta-Ulam-Tsingou Recurrence and Its Exact Dynamics,
Phys. Rev. X 8, 041017 (2018).

\bibitem{Yin2023} H. M. Yin and K. W. Chow, Fermi-Pasta-Ulam-Tsingou
recurrence and cascading mechanism for resonant three-wave interactions,
Phys. Rev. E 107, 064215 (2023).

\bibitem{Chowdury2025} A. Chowdury, B. J. Eggleton, and D. T. H. Tan,
Theoretical studies of modulation instability, Fermi-Pasta-Ulam recurrence
and pattern formation in an ultra-silicon-rich-nitride Bragg grating,
Nanophotonics 14, 2267 (2025).

\bibitem{Erkintalo2011} M. Erkintalo, K. Hammani, B. Kibler, C. Finot, N.
Akhmediev, J. M. Dudley, and G. Genty, Higher-Order Modulation Instability
in Nonlinear Fiber Optics, Phys. Rev. Lett. 107, 253901 (2011).

\bibitem{Liu2023a} L. Liu, W.-R. Sun, and B. A. Malomed, Formation of Rogue
Waves and Modulational Instability with Zero-Wavenumber Gain in
Multicomponent Systems with Coherent Coupling, Phys. Rev. Lett. 131, 093801
(2023).

\bibitem{Madimabe2023} E. B. Madimabe, C. B. Tabi, C. G. L. Tiofack, and T.
C. Kofan\'{e}, Modulational instability in vector exciton-polariton
condensates with photonic spin-orbit coupling, Phys. Rev. B 107, 184502
(2023).

\bibitem{Mossman2025} S. Mossman, S. I. Mistakidis, G. C. Katsimiga, A.
Romero-Ros, G. Biondini, P. Schmelcher, P. Engels, and P. G. Kevrekidis,
Nonlinear Stage of Modulational Instability in Repulsive Two-Component
Bose-Einstein Condensates, Phys. Rev. Lett. 135, 113401 (2025).

\bibitem{Chen2023} S.-C. Chen, C. Liu, and N. Akhmediev, Higher-order
modulation instability and multi-Akhmediev breathers of Manakov equations:
Frequency jumps over the stable gaps between the instability bands, Phys.
Rev. A 107, 063507 (2023).

\bibitem{Sun2024} Y. Sun et al., Multimode solitons in optical fibers:
a review, Photonics Research 12, 2581 (2024).

\bibitem{Lin2011} Y.-J. Lin, K. Jim\'{e}nez-Garc\'{\i}a, and I. B. Spielman,
Spin-orbit-coupled Bose-Einstein condensates, Nature 471, 83 (2011).

\bibitem{SOC-Review1}
I. B. Spielman. Light induced gauge fields for ultracold neutral atoms, Rev. Cold At. Mol. 1, 145 (2013).

\bibitem{SOC-Review2} N. Goldman, G. Juzeliunas, P. \"{O}hberg, and I. B.
Spielman, Light-induced gauge fields for ultracold atoms, Rep. Prog. Phys.
77, 126401 (2014).

\bibitem{SOC-Review3} H. Zhai, Degenerate quantum gases with spin--orbit
coupling: a review, Rep. Prog. Phys. 78, 026001 (2015).

\bibitem{fei2026} F.-Y. Liu and Q. Zhou, Periodic modulation on phase and
backgrounds for breathers and rogue waves in two-component Bose-Einstein
condensates, Chaos 36, 013125 (2026).

\bibitem{BenLi} H. Sakaguchi and B. Li, Vortex lattice solutions to the
Gross-Pitaevskii equation with spin-orbit coupling in optical lattices,
Phys. Rev. A 87, 015602 (2013).

\bibitem{HS} H. Sakaguchi, B. Li, and B. A. Malomed, Creation of
two-dimensional composite solitons in spin-orbit-coupled self-attractive
Bose-Einstein condensates in free space, Phys. Rev. E 89, 032920 (2014).

\bibitem{HP} Y.-C. Zhang, Z.-W. Zhou, B. A. Malomed, and H. Pu, Stable
solitons in three dimensional free space without the ground state:
Self-trapped Bose-Einstein condensates with spin-orbit coupling, Phys. Rev.
Lett. 115, 253902 (2015).

\bibitem{Sherman} H. Sakaguchi, B. Li, E. Ya. Sherman, and B. A. Malomed,
Composite solitons in two-dimensional spin-orbit coupled self-attractive
Bose-Einstein condensates in free space, Romanian Reports in Physics 70, 502
(2018).

\bibitem{Chisholm2026} C. S. Chisholm, S. Hirthe, V. B. Makhalov, R. Ramos,
R. Vatr\'{e}, J. Cabedo, A. Celi, and L. Tarruell, Probing
supersolidity through excitations in a spin-orbit-coupled Bose-Einstein
condensate, Science 391, 480--484 (2026).

\bibitem{Ding2024} C.-C. Ding, Q. Zhou, and B. A. Malomed,
Ultra-high-amplitude Peregrine solitons induced by helicoidal spin-orbit
coupling, Phys. Rev. Research 6, L032036 (2024).

\bibitem{Ding2025} C.-C. Ding, Q. Zhou, and B. A. Malomed, Beating stripe
solitons arising from helicoidal spin-orbit coupling in Bose-Einstein
condensates, Phys. Rev. E 111, 044203 (2025).

\bibitem{Ding2026} C.-C. Ding, Q. Zhou, and B. A. Malomed, Multi-pole
solitons and breathers with spatially periodic modulation induced by the
helicoidal spin-orbit coupling, Physica D 486, 135063 (2026).

\bibitem{Bhat2015} {I. A. Bhat, T. Mithun, B. A. Malomed, and K. Porsezian,
Modulational instability in binary spin-orbit-coupled Bose-Einstein
condensates, Phys. Rev. A 92, 063606 (2015).}

\bibitem{Mithun2019} {T. Mithun and K. Kasamatsu,
Modulation instability associated nonlinear dynamics of spin-orbit coupled
Bose-Einstein condensates, J. Phys. B: At. Mol. Opt. Phys. 52, 045301 (2019).}

\bibitem{Tabi2023} C. B. Tabi, E. Wamba, E. Nare, and T. C. Kofan\'{e},
Interplay between spin-orbit couplings and residual interatomic interactions
in the modulational instability of two-component Bose-Einstein condensates,
Phys. Rev. E 107, 044206 (2023).

\bibitem{Tabi2022a} C. B. Tabi, P. Otlaadisa, and T. C. Kofan\'{e},
Modulation instability of two-dimensional Bose-Einstein condensates with
helicoidal and a mixture of Rashba-Dresselhaus spin-orbit couplings, Physics
Letters A 449, 128334 (2022).

\bibitem{Otlaadisa2021} P. Otlaadisa, C. B. Tabi, and T. C. Kofan\'{e},
Modulation instability in helicoidal spin-orbit coupled open Bose-Einstein
condensates, Phys. Rev. E 103, 052206 (2021).

\bibitem{Bychkov1984} Y. A. Bychkov and E. I. Rashba, Oscillatory effects
and the magnetic susceptibility of carriers in inversion layers, J. Phys. C:
Solid State Phys. 17, 6039 (1984).

\bibitem{Dresselhaus1955} G. Dresselhaus, Spin-Orbit Coupling Effects in
Zinc Blende Structures, Phys. Rev. 100, 580 (1955).

\bibitem{Achilleos2013} V. Achilleos, D. J. Frantzeskakis, P. G. Kevrekidis,
and D. E. Pelinovsky, Matter-Wave Bright Solitons in Spin-Orbit Coupled
Bose-Einstein Condensates, Phys. Rev. Lett. 110, 264101 (2013).

\bibitem{Suret2024} {P. Suret, S. Randoux, A. Gelash, D. Agafontsev, B. Doyon,
and G. El, Soliton gas: Theory, numerics, and experiments, Phys. Rev. E 109,
061001 (2024).}

\bibitem{chong2022} C. Liu, S.-C. Chen, X. Yao, and N. Akhmediev, Modulation
Instability and Non-Degenerate Akhmediev Breathers of Manakov Equations,
Chinese Phys. Lett. 39, 094201 (2022).
\end{thebibliography}
\end{document}